\documentclass[useAMS,usenatbib]{mn2e}
\usepackage{epsfig}

\newcommand{\be}{\begin{equation}}
\newcommand{\ee}{\end{equation}}

\newcommand{\mnras}{MNRAS\,\,}

\newcommand{\aap}{A\& A\,\,}

\newcommand{\apj}{ApJ\,\,}

\def\Msun{\mbox{$M_\odot$}}
\def\Zsun{\mbox{$Z_{\odot}$}}
\def\lsim{\mathrel{\rlap{\lower3.5pt\hbox{\hskip0.5pt$\sim$}}
    \raise0.5pt\hbox{$<$}}}
\def\gsim{~\rlap{$>$}{\lower 1.0ex\hbox{$\sim$}}}

\def\tg{\mbox{$t_{gal}$}}
\def\tA{\mbox{$t_{AGN}$}}

\title[AGN Jet-induced Feedback in Galaxies.]{AGN Jet-induced Feedback in Galaxies. \\ II. Galaxy colours from a multicloud simulation}

\author[Tortora et al.]{\noindent
C.~Tortora$^{1}$\thanks{E-mail: ctortora@na.astro.it},
V.~Antonuccio-Delogu$^{3,2,1}$, S.~Kaviraj$^{2}$, J.~Silk$^{2}$,
A.D.~Romeo$^{1,4}$, \and U. Becciani$^{1}$
\\~\\
$^1$ INAF -- Osservatorio Astrofisico di Catania, Via S. Sofia 78,
I-95123 - Catania, ITALY \\
$^2$ Astrophysics, Department of Physics, University of Oxford,
Keble Road Ox1 3RH, Oxford, UNITED KINGDOM \\
$^3$ Institute for Theoretical Astrophysics, University of
Heidelberg, Albert-Ueberle-Str. 2, 69120 Heidelberg, GERMANY \\
$^4$ Universidad Andres Bello, Departamento de Ciencias Fisicas,
Av. Republica 252, Santiago, CHILE}

\begin{document}

\date{Accepted  Received }

\pagerange{\pageref{firstpage}--\pageref{lastpage}} \pubyear{2002}

\maketitle

\label{firstpage}

\begin{abstract}
We study the feedback from an AGN on stellar formation within its
host galaxy, mainly using one high resolution numerical simulation
of the jet propagation within the interstellar medium of an
early-type galaxy. In particular, we show that in a realistic
simulation where the jet propagates into a two-phase ISM, star
formation can initially be slightly enhanced and then, on
timescales of few million years, rapidly quenched, as a
consequence both of the high temperatures attained and of the
reduction of cloud mass (mainly due to Kelvin-Helmholtz
instabilities). We then introduce a model of (prevalently) {\em
negative} AGN feedback, where an exponentially declining star
formation is quenched, on a very short time scale, at a time \tA,
due to AGN feedback. Using the \cite{BC03} population synthesis
model and our star formation history, we predict galaxy colours
from this model and match them to a sample of nearby early-type
galaxies showing signs of recent episodes of star formation
(Kaviraj et al. 2007). We find that the quantity $\tg - \tA$,
where \tg\ is the galaxy age, is an excellent indicator of the
presence of feedback processes, and peaks significantly around
$\tg - \tA \approx 0.85 \, \rm Gyr$ for our sample, consistent
with feedback from recent energy injection by AGNs in relatively
bright ($M_{B} \lsim -19$) and massive nearby early-type galaxies.
Galaxies that have experienced this recent feedback show an
enhancement of 3 magnitudes in $NUV(GALEX)-g$, with respect to the
unperturbed, no-feedback evolution. Hence they can be easily
identified in large combined near UV-optical surveys.
\end{abstract}


\section{Introduction}

Active galactic nuclei (AGN) have been advocated in recent years
as sources able to influence the evolutionary history of stellar
populations within their host galaxies. They are ubiquitous and
lie in the cores of a wide variety of galaxies, possibly
suppressing, or possibly on occasion speeding up, star formation
(SF). Massive and bright early-type galaxies (ETGs) have been
shown to host very old stellar populations where SF stopped early,
while fainter ETGs show lower ages and more protracted SF phases
(\citealt{Thomas05}, \citealt{deLucia06}). This is the
so-called downsizing scenario, which is predicted by observations
and generated in some recent simulations (\citealt{Cowie96},
\citealt{Borch06}, \citealt{Bundy06}, \citealt{deLucia06},
\citealt{Trager00}, \citealt{Thomas05}, \citealt{Nelan05},
\citealt{Tortora09}). However, only very recently
semi-analytic simulations have been able to reproduce some of the
key observations (\citealt{Cattaneo06}, \citealt{deLucia06}).

\cite{KC98} predicted results that contrast with the downsizing
scenario. The main reason for this discrepancy may be that
supernova feedback in massive galaxies is not sufficient to quench
SF (\citealt{DS86}, \citealt{Benson03}). Feedback from AGN is the
additional source able to heat the cold gas that fuels SF in
galaxies. It is such effects that allow us to match the evolution
of massive galaxies predicted from simulations with results from
observations (e.g., \citealt{Kaviraj05}, \citealt{Silk05},
\citealt{Bower06}, \citealt{Croton06}, \citealt{deLucia06}).
However, these studies did not account for the downsizing
phenomenon.

Feedback from AGN plays a major role during different phases of
galaxy evolution. Higher activity is observed at high redshifts.
At these early phases of galaxy evolution, during a major merger,
the gas accreted by central supermassive black holes is ejected
within the host galaxy, and inhibits SF (\citealt{DiMatteo05},
\citealt{Springel05, Springel05b}). At more recent epochs, the
effect of AGN feedback is weaker, but still sufficient to quench
SF, particularly near the central regions (\citealt{Croton06},
\citealt{Bower06}, \citealt{Schawinski06}, 2008). Evidence for the
presence of AGNs is evident in many previous  studies. In
particular, \cite{K2003} collected more than 20000 narrow-line
AGNs from SDSS at $z \lsim 0.3$. Type-II AGN are found within host
galaxies having structural properties similar to normal ETGs, and
reside almost exclusively in massive galaxies with $M_{\star} >
10^{10}\, \rm M_{\odot}$. Many of these galaxies have experienced
a recent burst of SF or show evidence of ongoing SF, thus
indicating that a black hole and prominent source of fuel supply
are necessary ingredients for AGN feedback, but no evidence
of SF quenching is found (\citealt{K2003}). While the existence
of a black hole within massively star forming galaxies is a rarity
in our cosmological neighbors, the situation is different at
higher redshifts, where the AGN frequency is higher
(\citealt{Hasinger05}) and the feedback could be effective in
producing red galaxies (\citealt{Benson03}, \citealt{Cattaneo06},
\citealt{Schawinski08}).

More recently, a series of analyses have been devoted to the
detection of recent star formation (RSF) within the last Gyrs of
galaxy lifetimes. Ultraviolet data allowed the detection of small
fractions of young stars, allowing one to trace this residual SF
in ETGs both in the local universe (\citealt{Yi05},
\citealt{Kaviraj07}) and at high redshift (\citealt{Kaviraj08}).
The link between SF and AGNs is strengthened by other analyses,
where AGNs have been found to be responsible for the migration
of star-forming galaxies, lying within the blue cloud, to almost quiescent
located in the red sequence (\citealt{Schawinski06}, 2007, 2008,
\citealt{Bildfell08}). In \cite{Kaviraj07b}, the recent
feedback effect in E+A galaxies is analyzed and the bimodal
correlation of the quenching efficiency with mass and luminosity
is interpreted as due to negative feedback from supernovae or
AGNs.

{The physics of the SF-AGN connection is still poorly
understood, and the empirical models of SF quenching by AGN used
in different works (\citealt{Granato01, Granato04},
\citealt{Martin07}), although reasonable, do not yet have a
solid motivation. Thus, a primary goal should be to analyze this
phenomenology by means of hydrodynamical simulations,
modelling the propagation of jets produced by AGNs within a
gaseous medium (\citealt{Scheuer74}, \citealt{Falle91}) and their
interaction with an inhomogeneous interstellar medium
(\citealt{Saxton05}, \citealt{KA07}, \citealt{SB07}). In a
previous paper (\citealt{AS08}, paper I hereafter) we used an
Adaptive Mesh Refinement (AMR) code to follow the evolution of the
cocoon produced by the jet propagating in the ISM/IGM: we have
analyzed the thermodynamic evolution of a cloud embedded within
the cocoon and  seen how its SF is modified. We found the SF  to
be quenched on time-scales of $10^5-10^6 \, \rm yrs$.

In the present paper, we deduce a Star Formation History
(hereafter SFH) directly from a numerical simulation of the
interaction of an AGN with the multiphase ISM of its host galaxy.
In principle, a realistic simulation should at least encompass a
range of scales spanning more than 5 orders of magnitudes in the
spatial coordinates, down to single star-formation scales, and should
include a very wide range of heating and cooling phenomena. In
this work we have restricted ourselves to consider only the
\emph{mechanical} feedback from the AGN, and its effect on the
thermodynamic state of the ISM, thus neglecting the
\emph{radiative} feedback. We have modelled the ISM as a
two-phase, multicloud system, and we have followed the
thermodynamic evolution of both phases, particularly of the cold,
star-forming phase. We introduce a feedback model where {\em
negative feedback}, i.e. the suppression of SF within the volume
affected by the jet, plays a dominant role. Using an empirical
prescription to take into account the effect of cocoon on SF
within cold clouds, we present a scenario where SF in unperturbed
ETGs is described by an exponential SF law, corresponding to an
early starburst followed by a slow decline. This evolution,
starting at some time \tA\ is quenched by a feedback process, and
we follow the evolution of galaxy colors, determining the free
parameters such as galaxy age \tg\ and \tA\ by comparing with a
previously studied sample (\citealt{Kaviraj07}).

For the first time, the amplitude and typical timescales of
negative AGN feedback are directly derived from a simulation
designed to address this phenomenology.

The plan of the paper is as follows. In Sect. \ref{sec:sim_setup}
we discuss the simulation set-up, while in Sect.
\ref{sec:cocoon_prop} we describe the propagation of the cocoon and
its influence on SF. In order to compare predicted results with
observations, we use single burst population models from
\cite{BC03}, suitably convolved with the SF rate predicted by our
simulation: we describe our procedure and predicted synthetic
colours in Sect. \ref{sec:colours}. Finally, Sect.
\ref{sec:Observations} is devoted to a comparison with
observations, and conclusions are detailed in Sect.
\ref{sec:conclusions}.

Whenever required, we will adopt the  3-year WMAP ``{\it
concordance}'' model: h=0.74,\, $\Omega_{m} = 0.234,\,
\Omega_{b}h^{2} = 0.0223$, corresponding to a Universe age of
$t_{\rm Univ} \sim 13.6 \, \rm Gyr$.

\begin{figure*}
\epsfig{file=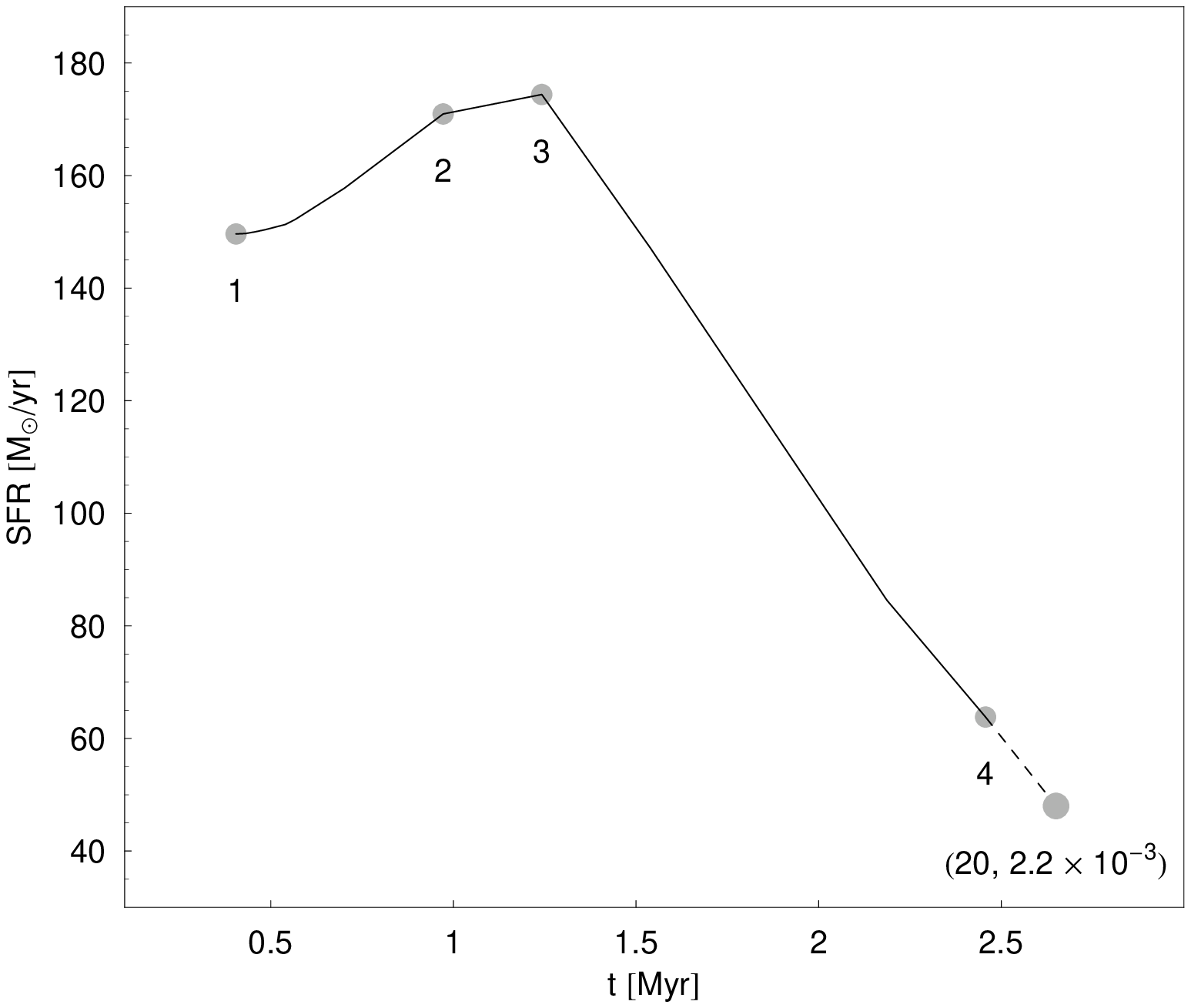, width=0.65\textwidth}\\
\vspace{0.5cm} \epsfig{file= 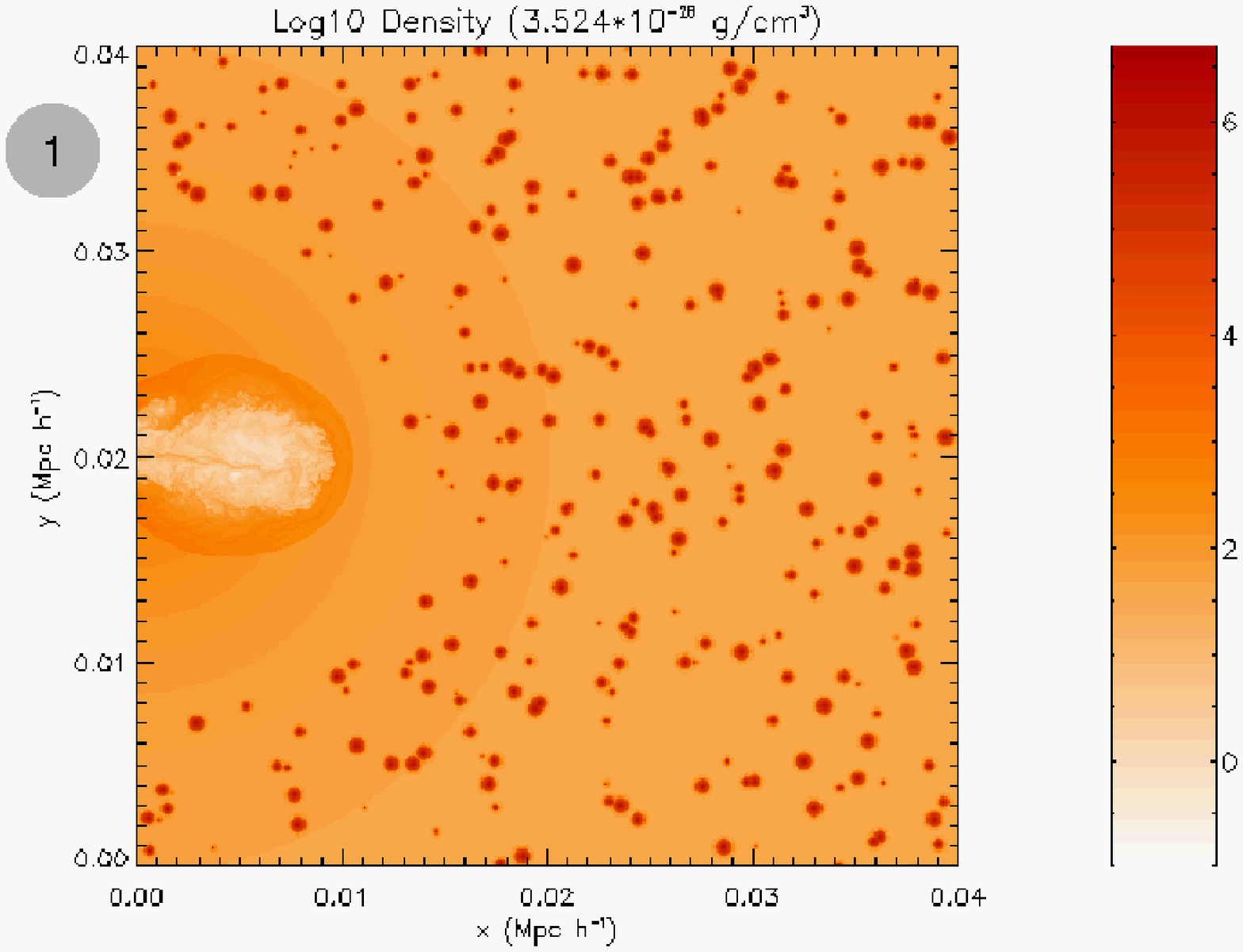,
width=0.38\textwidth} \hspace{0.5cm}
\epsfig{file= 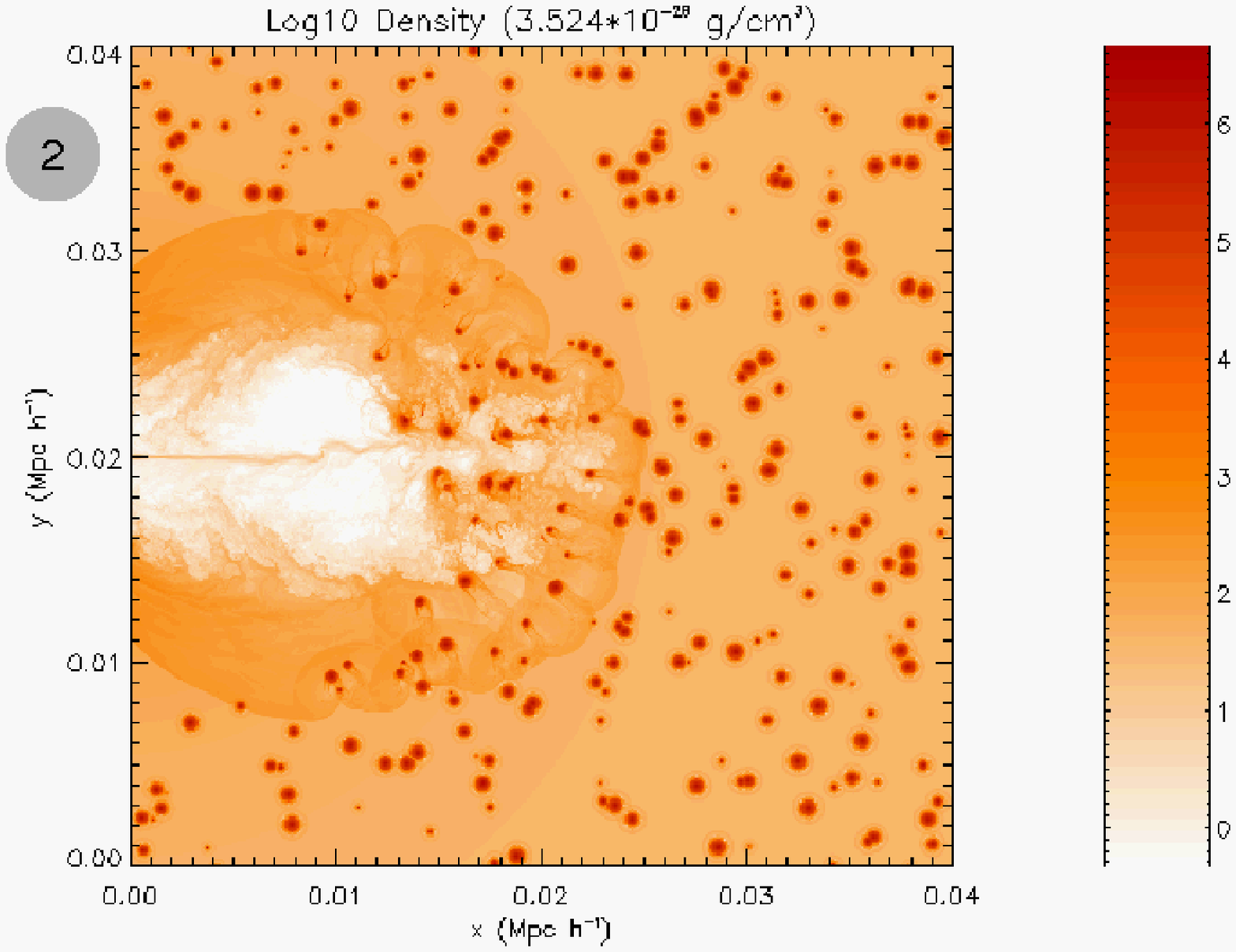, width=0.38\textwidth}\\
\vspace{0.5cm} \epsfig{file= 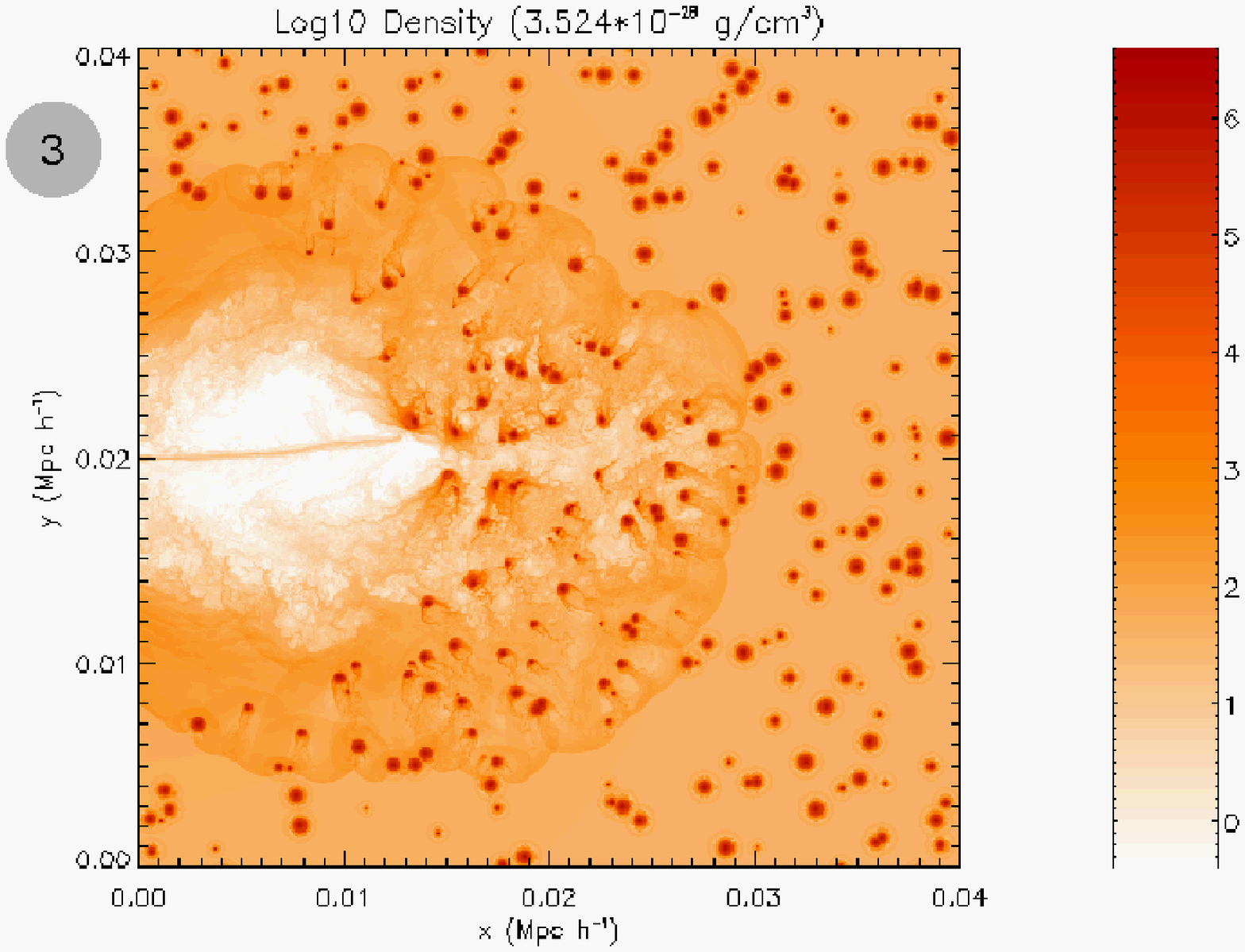,
width=0.38\textwidth}\hspace{0.5cm} \epsfig{file=
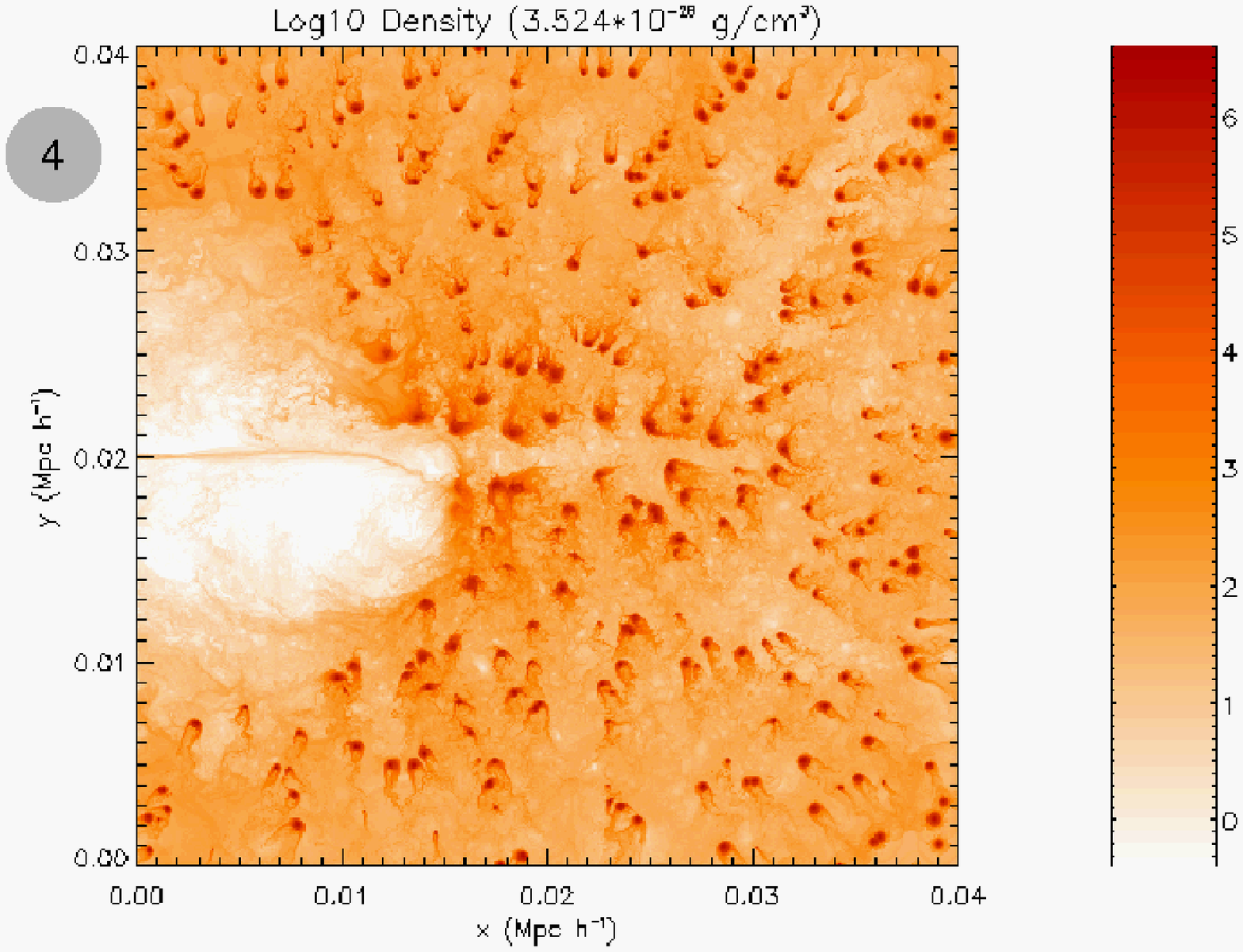, width=0.38\textwidth} \caption{Temporal
evolution of the star formation rate, for different snapshots of
the simulation. In the upper panel we show the SF as a function of
time (measured in $\Msun\ / yr$ and $\rm Myr$ respectively),
marking with numbers $1,\, 2,\, 3,\, 4$ the SFs at the times $t
\approx 0.4, \, 1.0, \,1.2, \,2.5 \, \rm Myr$ respectively. In the
four bottom panels we show the propagation of the cocoon within
the ISM and the effect on clouds (corresponding to the numbers in
the top panel). SF rate is obtained using the Schmidt-Kennicutt
law ($n=1.4$) and collecting the clouds within a radius of $27 \,
\rm kpc$. At very late epochs ($t \sim 20 \, \rm Myrs$) the SFR
declines to a very small value, as seen in the top panel. Note
that our simulation maps only one-half of the galaxy, thus,
assuming symmetry, we multiply our SF by a factor 2. }
\label{fig:SFRmed_vs_t_snaphots}
\end{figure*}

\section{Simulation setup}\label{sec:sim_setup}

To perform the simulation, we used FLASH v.2.5
(\citealt{Fryxell00}), a parallel, Adaptive Mesh Refinement (AMR)
code, which implements a second order, shock-capturing PPM solver.
The modular structure of FLASH allows the inclusion of various
physical effects, including a source of external heating,
radiative cooling and thermal conduction. In our simulation, we
include radiative cooling, using the standard cooling function
from \cite{SD93}, conveniently extended towards higher
temperatures, $T> 10^{7} \, \rm K$, which are attained within
the cocoon generated in this simulation (see App. A in Paper I).
We also take into account gravity, but we neglect thermal
conduction and large-scale, ordered magnetic fields\footnote{Even an
initially  weak, small-scale, tangled magnetic field would be
amplified to a level which inhibits thermal conduction, by 2-3
orders of magnitude w.r.t. the classical Spitzer values.}.

The jet is modelled as a one-component fluid, characterized by a
density $\rho_{j}$ which is a fraction $\epsilon_{j}$ of the
initial density of the interstellar medium. In order to prevent
numerical instabilities at the jet/ISM injection interface, we use
a steep, continuous and differentiable transverse velocity and
density profile, as we did before (paper I; \citealt{Perucho04},
2005): \be v_{x,j} =
\frac{V_{j}}{\cosh\left\{\left(\frac{y-y_{j}}{d_{j}}\right)^{\alpha_{j}}\right\}}
\label{eq:vinjet} \ee \be n_{j} = n_{env} - \frac{\left(n_{env} -
n_{j}\right)}{\cosh\left\{\left(\frac{y-y_{j}}{d_{j}}\right)^{\alpha_{j}}\right\}}
\label{eq:rhojet} \ee where: $\alpha_{j}=10$ is an exponent which
determines the steepness of the injection profile, $n_{j},
n_{env}$ denote the jet and environment electron number densities
and the scalelength $d_{j}$ characterizes the width of the jet.
The power injected by the jet is then given by $P_{j} = \beta 2\pi
d_{j}^{2}\rho_{j}V_{j}^{3}$, with $\beta \simeq 0.7158$. In this
simulation we set: $P_{j} = 10^{46} \, \rm erg \, s^{-1}$.

We model the environment, where the jet propagates, as a 2-phase
interstellar medium, comprising a hot, diffuse, low-density
component and a cold, clumped system of high density clouds in
pressure equilibrium with the diffuse component. The warm phase is
characterized by a density profile $\rho_{env}(r)$ and a constant
temperature $T_{env}$. We assume that the diffuse gas is embedded
within a dark matter halo, the latter being described by a NFW
density profile. This DM halo is chosen to have a total mass
$M_{h} = 5\times 10^{11}\, {\rm M}_{\sun}$, concentration $c =
10.2$ and a scalelength $l_{h} = 206\, h^{-1}\,$ kpc. The ISM gas
is assumed to be a fraction $m_{g} = \Omega_{gas}/\Omega_{DM}
\approx 0.212 $ of the DM, and to be distributed in hydrostatic
equilibrium within this DM halo, following the prescription given
in Appendix C of \citet{2006ApJ...647..910H}. Each of the clouds
in the cold component is modelled as a truncated isothermal sphere
(TIS: \citealt{1999MNRAS.307..203S},
\citealt{2001MNRAS.325..468I}), because this model seems to
adequately reproduce  the properties of clouds formed in
simulations of a thermally unstable ISM. TIS spheres posses a
finite radius, and are characterized by two parameters, which we
assume to be the mass $M_{cl}$ and a typical radius $r_{cl}$. They
are exact solutions of the equilibrium equations for isothermal
spheres confined by an external pressure. We distribute 300 clouds
within the simulation volume, using a mass spectrum previously
derived from numerical simulations
(\citealt{2005ApJ...630..689B}). Summarizing, the simulation is
specified by 10 parameters: three of these describe the DM halo
($M_{h}, c, l_{h}$), two the diffuse phase of the ISM ($m_{g},
T_{env}$), two the mass distribution within the cold component of
the ISM ($M_{cl}, r_{cl}$), and the last three describe the jet
($n_{j}, d_{j}, P_{j}$).

\begin{table}
\centering \caption{Parameters of the simulation. The halo mass is
in units of $M_{\sun}$, distances are measured in kpc $h^{-1}$,
$n_{env}, n_{j}$ are, respectively, the central gas electron and jet's densities, in units of $\rm cm^{-3}$, $T_{env}$ is in Kelvin,
$d_{j}$ and $P_{j}$ are the jet width and power, respectively. The
latter is expressed in cgs units (ergs$\cdot$ sec.}
\begin{tabular}{lc}
\hline \hline
Parameter & Adopted value \\
\hline
$M_{h}$ & $5\times 10^{11}$  \\
$c$ & 10.2  \\
$l_{h}$ & 206  \\
$m_{g}$ & $1.06\times 10^{11}$ \\
$n_{env}$ & $10^{2}$\\
$T_{env}$ & $10^{7}$\\
$n_{j}$ & $1$ \\
$d_{j}$ & 10\\
$P_{j}$ & $10^{46}$ \\
\hline \hline
\end{tabular}\label{tab:cl}
\end{table}

\begin{figure}
\psfig{file= 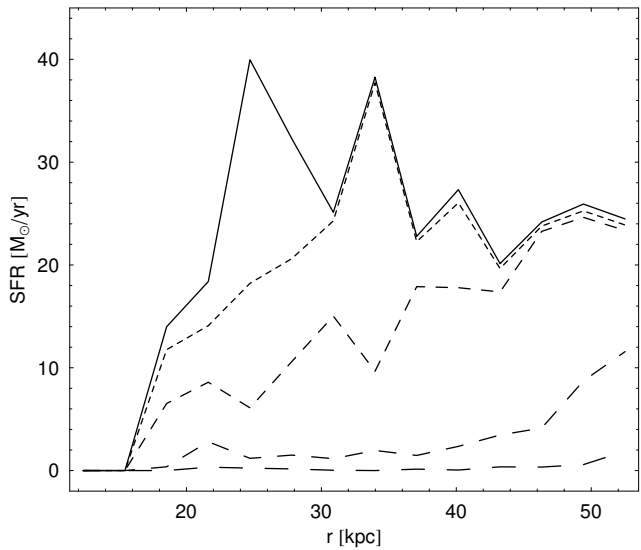, width=0.48\textwidth}\caption{Star formation
rate in 14 circular, equally-spaced shells. The results for different times are
shown, from continuous to long dashed lines spanning $t = 0.4, \,
1, \,1.6, \,2.5, \, 3.2 \, \rm Myr$.} \label{fig:SFRmed_vs_r}
\end{figure}

We choose a simulation box having a size: $L_{box} = 40\, h^{-1}\,
\rm kpc$, so that the jet will diffuse through it at the end of
the simulation. The spatial resolution attained is a function
of the maximum refinement level and of the structure of the
code. For a block-structured AMR code like FLASH, where each block
is composed by ${\rm n}_{x}\times{\rm n}_{y}$ cells, the maximum
resolution along each direction is given by $L_{box}/({\rm
n}_{x}2^{l})$, where $l$ is the maximum refinement level. In this
simulation: ${\rm n}_{x} = {\rm n}_{y} = 8\,$ and $l=6$, thus the
minimum resolved scale is $78.125\,$ pc. Note that we are
performing a 2D simulation, but we do not impose any special
symmetry.

We assume that the duty cycle of the jet is active for a time of
$4\times 10^{6}$ yrs, as is  typical for jets having such power
(\citealt{2008MNRAS.388..625S}), and its power declines
linearly with time since this epoch until it switches off at $t~\approx
2\times 10^{7}$ yrs. As for the star formation rate, we assume
that the clouds are converting gas into stars with a rate
specified by the Schmidt-Kennicutt (SK) law (\citealt{Schmidt59},
1963, \citealt{Kennicutt98}): $SFR = \dot{\Sigma} = A \Sigma^{n}$,
where $A=(2.5 \pm 0.17) \times 10^{-4} \, \rm \Msun \, yr^{-1} \,
kpc^{-2}$ and $n=1.4 \pm 0.15$. We further require that SF is
ongoing only within regions having a mass larger than the
Bonnor-Ebert mass (\citealt{Ebert55}, \citealt{Bonnor56}), and
temperature less than a specified upper limit, i.e. $T \leq 1.2
\times 10^{4} \, \rm K$. These are rather conservative
constraints, which tend to underestimate the magnitude of the
negative feedback.

Our initial setup is very different from that of the very
recent simulation by \cite{SB07}, particularly in one very
important point: while we model a population of pressure-confined
clouds \emph{embedded} within the ISM, \cite{SB07} put a turbulent
disk around the jet's source, embedding the AGN within it, as they are
more interested in addressing problems related to the evolution of
GPS and CSS radio sources. Also, note that their simulation box is
much smaller than ours (1 kpc), and the temporal scale
is also very different ($\approx 10^{5}\, \rm yrs$  in their
simulation, compared to $2\times 10^{7}\, \rm yrs$ in the present
work). Moreover, one may notice that we have neglected the
velocity field of the inhomogeneous component: in fact, as already
noted by \cite{SB07}, for the spatial and temporal scales of
interest, its inclusion has little effect on the evolution of
the clouds and of their SF. The typical turbulent velocities of
clouds are in the range $10-70 \, {\rm km \, s}^{-1}$. Thus, on
typical timescales $ \lsim 2 \times 10^{7}\, \rm yrs$ in our
simulation, these clouds have displacements $l \lsim 0.2 - 1.4\,
\rm kpc$, smaller than the \emph{average} distance among clouds
(see Fig. \ref{fig:SFRmed_vs_t_snaphots}).

\section{Propagation of the jet}\label{sec:cocoon_prop}

Soon after the jet enters the ISM, a low-density region, the
cocoon, is generated. The details of the propagation of the jet
within the ISM, and the properties of the cocoon, have been
extensively studied in the past (\citealt{Scheuer74},
\citealt{Falle91}). Only recently, however, numerical simulations
have been used to study in more detail the effect which this
interaction has on the inhomogeneous component of the ISM, where
stars are forming (e.g., \citealt{Saxton05}, \citealt{KA07}). In
Paper I, we extensively analyzed properties of cocoon and the
interaction with a single cloud that is forming stars. In the
following we will make few comments about cocoon propagation,
concentrating rather on the effects on our multicloud system.

\subsection{Quenching of star formation}\label{sec:SF}

The global evolution can be seen in
Fig.~\ref{fig:SFRmed_vs_t_snaphots}, it is possible to distinguish
two main phases, corresponding to the evolution of the cocoon: an
{\it active} and {\it passive} phase. While the jet is active, the
cocoon is fed and it expands almost self-similarly, while, when  it is
switched off, the cocoon diffuses and affects a larger fraction of
the ISM.

Initially (see snapshots 1 and 2 in Fig.
\ref{fig:SFRmed_vs_t_snaphots}), the interaction mostly takes
place through the weak shock at the interface between the cocoon
and the ISM. This shock compresses and heats up the clouds, having
two opposite effects on SF: compression tends to increase SF, but
the larger temperature also increases the critical mass for
gravitational collapse, thus decreasing the volume fraction of the
clouds which can actively form stars. Globally, we observe an
initial enhancement of SF (i.e., we observe a {\it positive
feedback}). The effects of the two concomitant processes on clouds
can be seen in Fig.~\ref{fig:SFRmed_vs_r}, where we build the
shell-SFs by adding the SF of clouds lying in a concentric annulus
of mean radius $r$ and plot them as a function of radius for
different time steps. Here, few episodes of positive feedback are
evident in the clouds which are nearer to the cocoon. Later ( see
snapshots 3 and 4 in Fig. \ref{fig:SFRmed_vs_t_snaphots}), the
cocoon propagates within the medium inducing a general increase of
the temperature, which heats up the clouds and decreases their
density, and finally the outer regions of the clouds are stripped
due to Kelvin-Helmholtz instabilities, as we studied in more
detail in Paper I. These effects tend to reduce the mass of the
clouds, decreasing SF, and eventually  drastically suppressing it
on a time-scale of $2-3 \times 10^{6} \, \rm yr$ ({\it negative
feedback}). Obviously, the more external regions in the galaxy are
affected later by the disruptive effect of the cocoon, being
almost unperturbed until $t \sim 10^{6} \, \rm yr$ and $t= 1.6
\times 10^{6} \, \rm yr$, respectively at $r \sim 30$ and $\sim 45
\, \rm kpc$ (see Fig. \ref{fig:SFRmed_vs_r}). As we see from this
simulation, approximately on a timescale over which the cocoon
diffuses, the clouds are destroyed: this is then the typical
time-scale over which SF will be inhibited. Obviously, one must
observe that this timescale is not exactly coincident with the
duty cycle of the jet.

\begin{figure}
\centering \psfig{file=
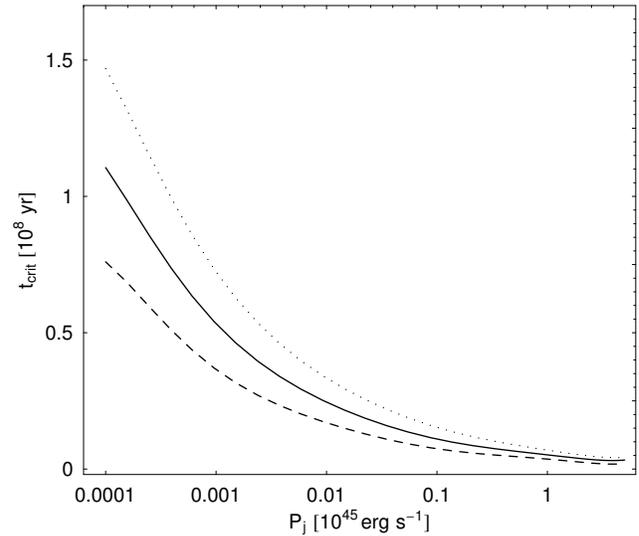,width=0.47\textwidth}\caption{Timescale for
feedback. The three curves refer to the same density profile,
having $\rho_{0}=1.7\times 10^{-25} \, \rm g \, cm^{-3}$,
$\beta=2$, and 3 values for $a_{0} = 0.8, \, 1, \,1.2\, \, \rm kpc
\, h^{-1}$ (dashed, continuous and dotted curves, respectively).
The strong dependence on $a_{0}$ is simply a consequence of the
high exponent with which it appears in the expression for
$t_{0}$.} \label{fig:tscale}
\end{figure}

\begin{figure*}
\psfig{file= 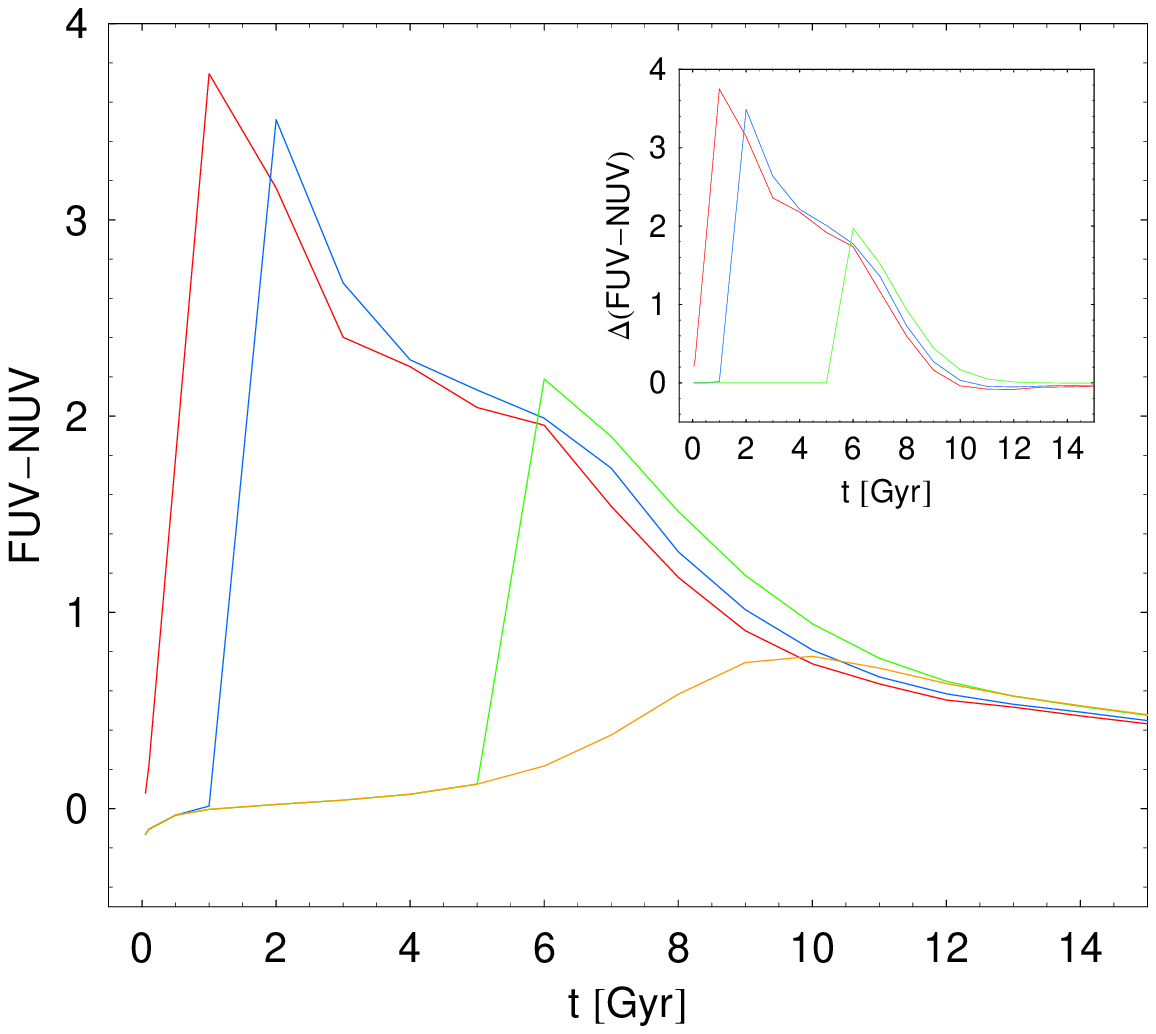,
width=0.32\textwidth}\hspace{0.2cm}\psfig{file= 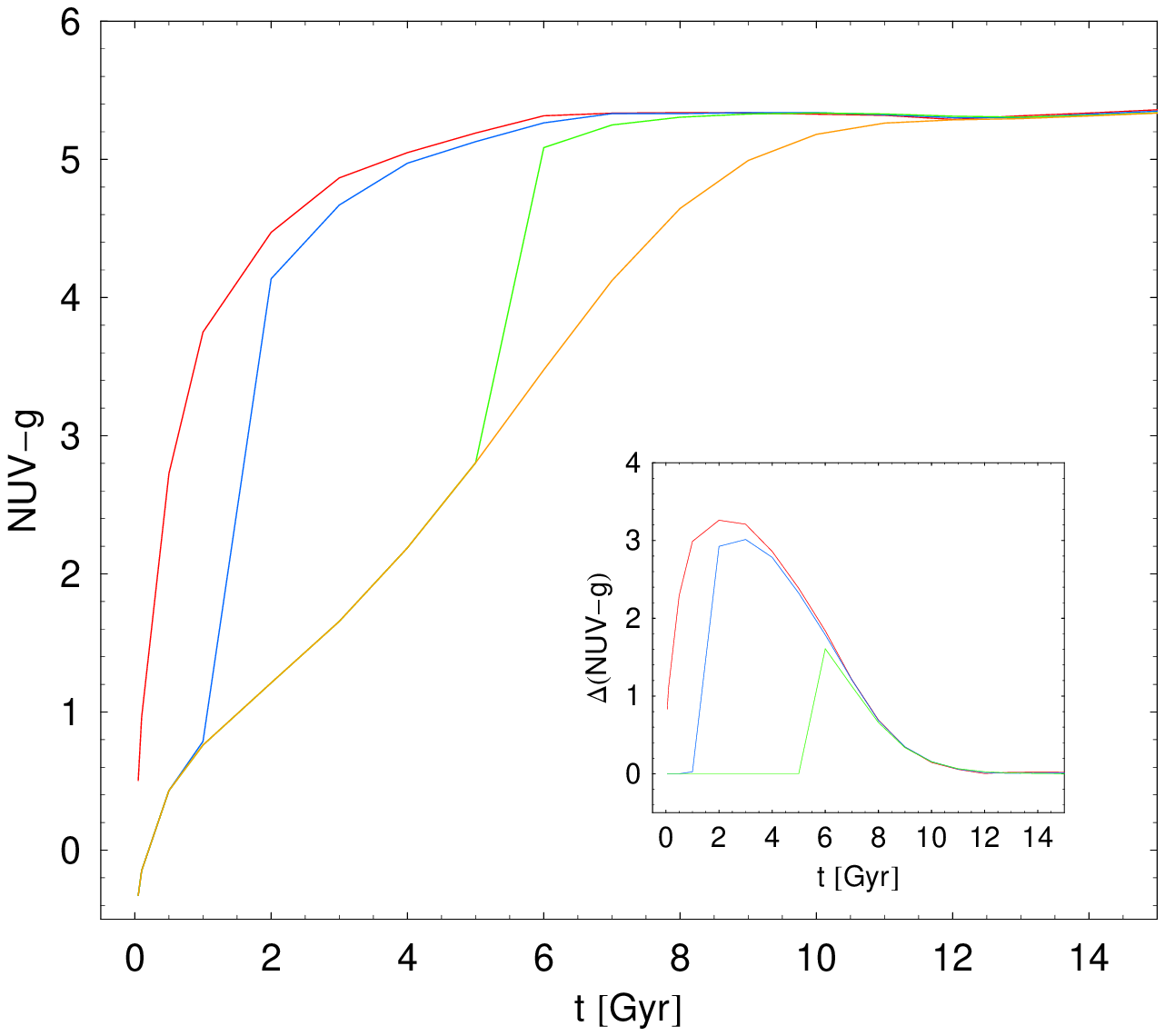,
width=0.32\textwidth}\hspace{0.2cm} \psfig{file=
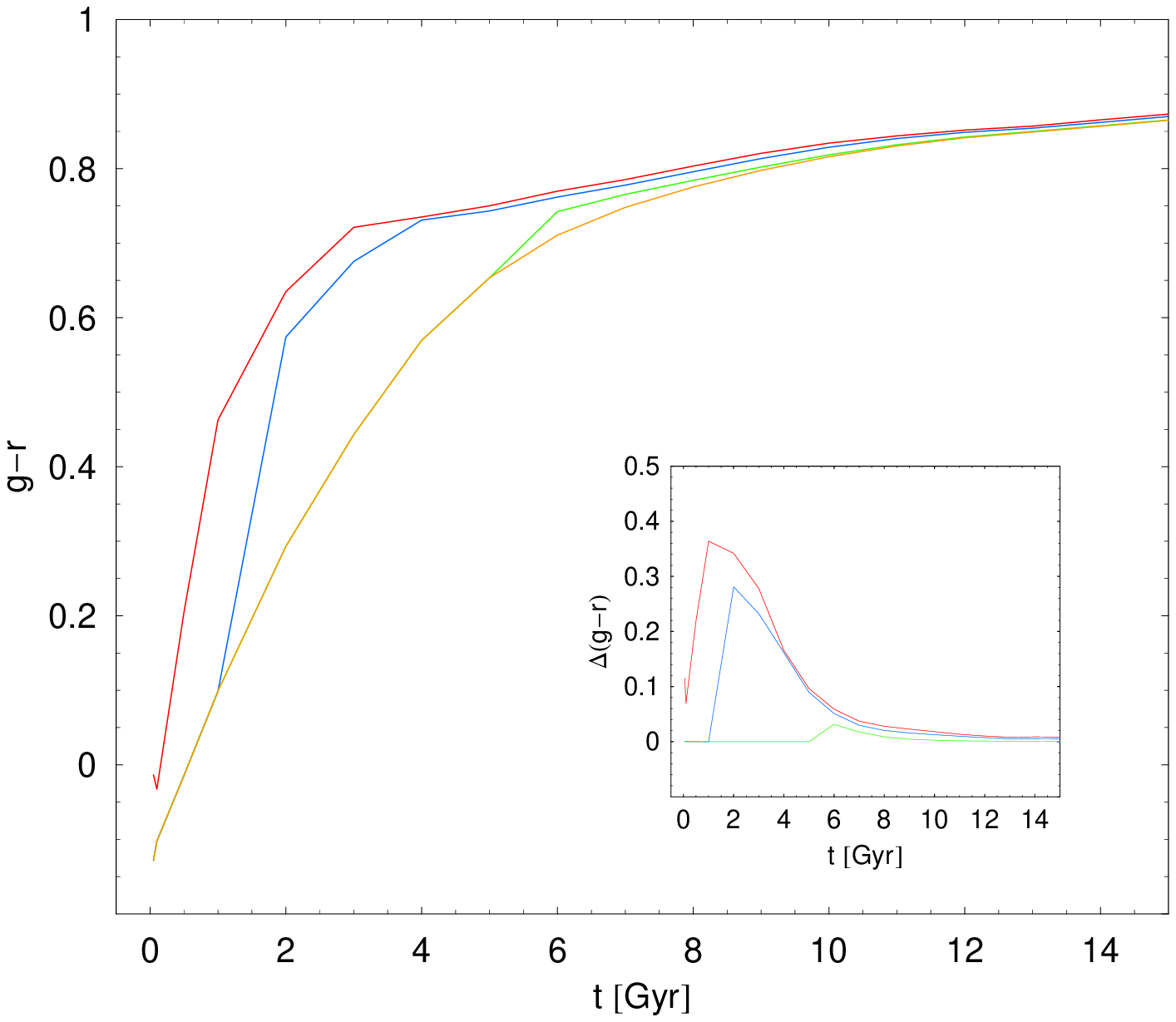, width=0.32\textwidth}\caption{Estimated
synthetic colours FUV-NUV, NUV-g and g-r (in AB system) as a
function of galaxy age (in units of $\rm Gyr$) for $Z=\Zsun$. Red,
blue, green and orange curves correspond respectively to an
unperturbed exponential SF with $\tau = 0$, an exponential SF with
$\tau = 1 \, \rm Gyr$ and $\tA =1 \, \rm Gyr$, an exponential SF
with $\tau = 1 \, \rm Gyr$ and $\tA =5 \, \rm Gyr$ and an
unperturbed exponential SF with $\tau = 1 \, \rm Gyr$. In the
inset panels we show the difference of colours with respect to the
reference one with an unperturbed exponential SF with $\tau = 1 \,
\rm Gyr$ as a function of galaxy age; the colour code is the same
as in the plot of colours vs age.} \label{fig:col_vs_t}
\end{figure*}

Following Paper I, we restricted our simulation to a single
episode of jet injection. Any subsequent event of jet injection
would have little influence on SF, since the jet would propagate
through a high temperature ($T \sim 10^{8}-10^{11} \, \rm K$) and
low density ($n_{e} \sim 10^{-2}-10^{-1} \, \rm cm^{-3}$)
environment. As already noted in \cite{Inuoe01} and in Paper I, after
the injection of the jet, the cooling time of the diffuse ISM
exceeds the dynamical time by a factor $6.5 \times 10^{2} - 3
\times 10^{5}$; thus the heated gas is not able to cool, quenching
the second emitted jet. The extent of this region crucially
depends on the physical parameters of the medium and on $P_{j}$.
The case  we present in this paper is that of a very powerful jet,
so at the end all the gas within the affected region is influenced by
the expansion of the cocoon. For less powerful jets, it is
reasonable to expect that the extent of the quenched SF region
will be much smaller.

We should however remark that this scenario could not be
exhaustive of all the possibilities. In a recent paper,
\cite{KA07} studied the evolution of clouds lying very near to a jet.
The clouds are destroyed by the Kelvin-Helmoltz instability
produced by the jet, but the authors notice that a fraction of
these clouds, under the effect of thermal instabilities, evolve
into filamentary structures which are characterized by high
densities. These filaments could then still host SF, although it
is not clear the \emph{quantitative} relevance of this phenomenon
to global SF within the host galaxy.

In \cite{SB07} the mechanical energy of the jet is mostly
dissipated when the jet tries to percolate through the disk, while
when the jet has already formed a cocoon, as in our simulation,
its energy feeds the turbulence within the cocoon. On the large
scales probed by our simulation, the turbulence within the cocoon
is responsible for the final destruction of the clouds. When the
jet is switched off this turbulence will decay but, as already
noticed by \citet{Inuoe01}, often the cooling time of the very hot
plasma is too long to allow the formation of cold clouds. Other
external episodes like minor interactions between clouds could
however fuel again some cold gas, and strip the hot gas, thus
favoring the subsequent development of embedded cold clouds.

\begin{figure*}
\psfig{file= 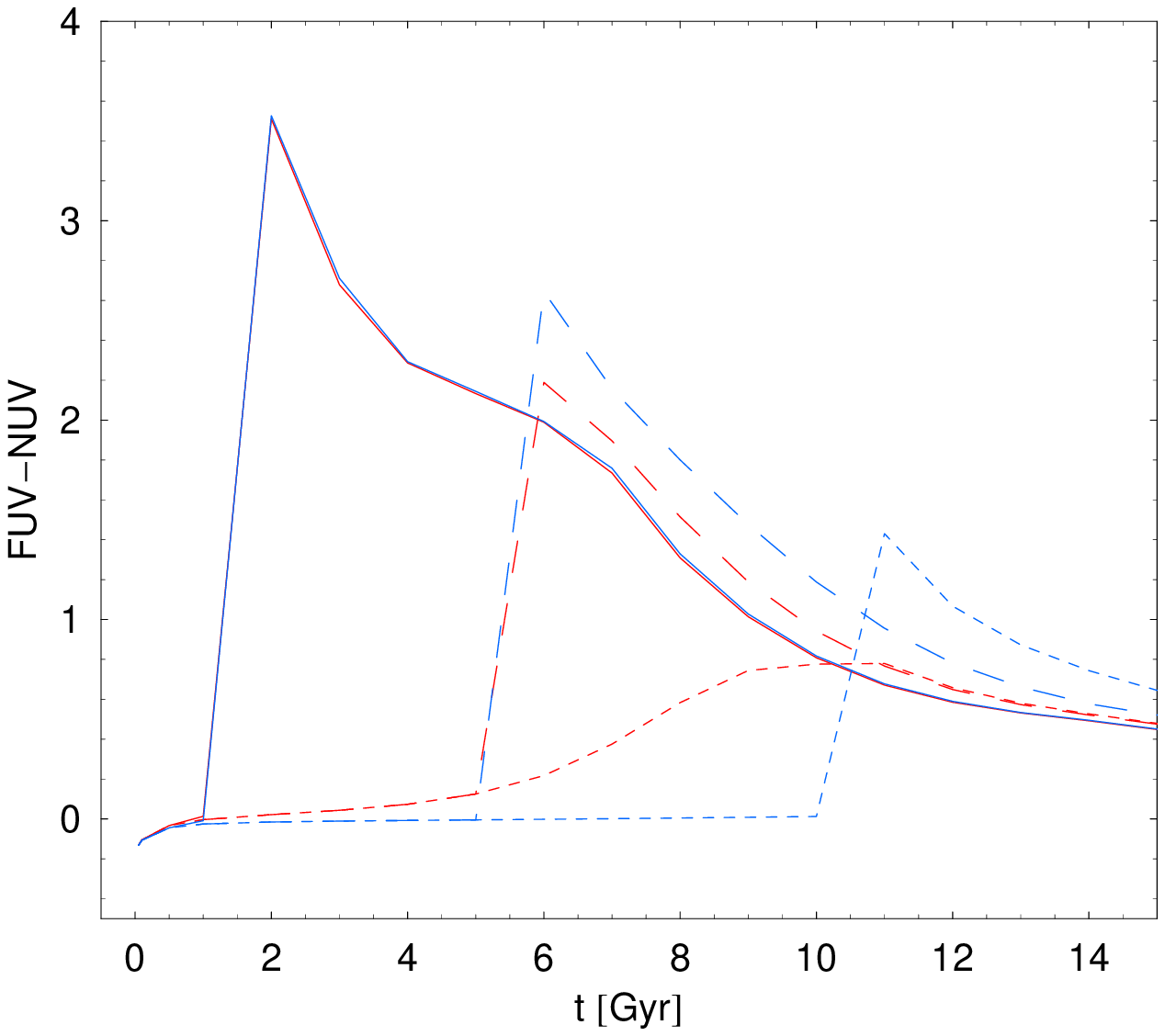,
width=0.32\textwidth}\hspace{0.2cm}\psfig{file= 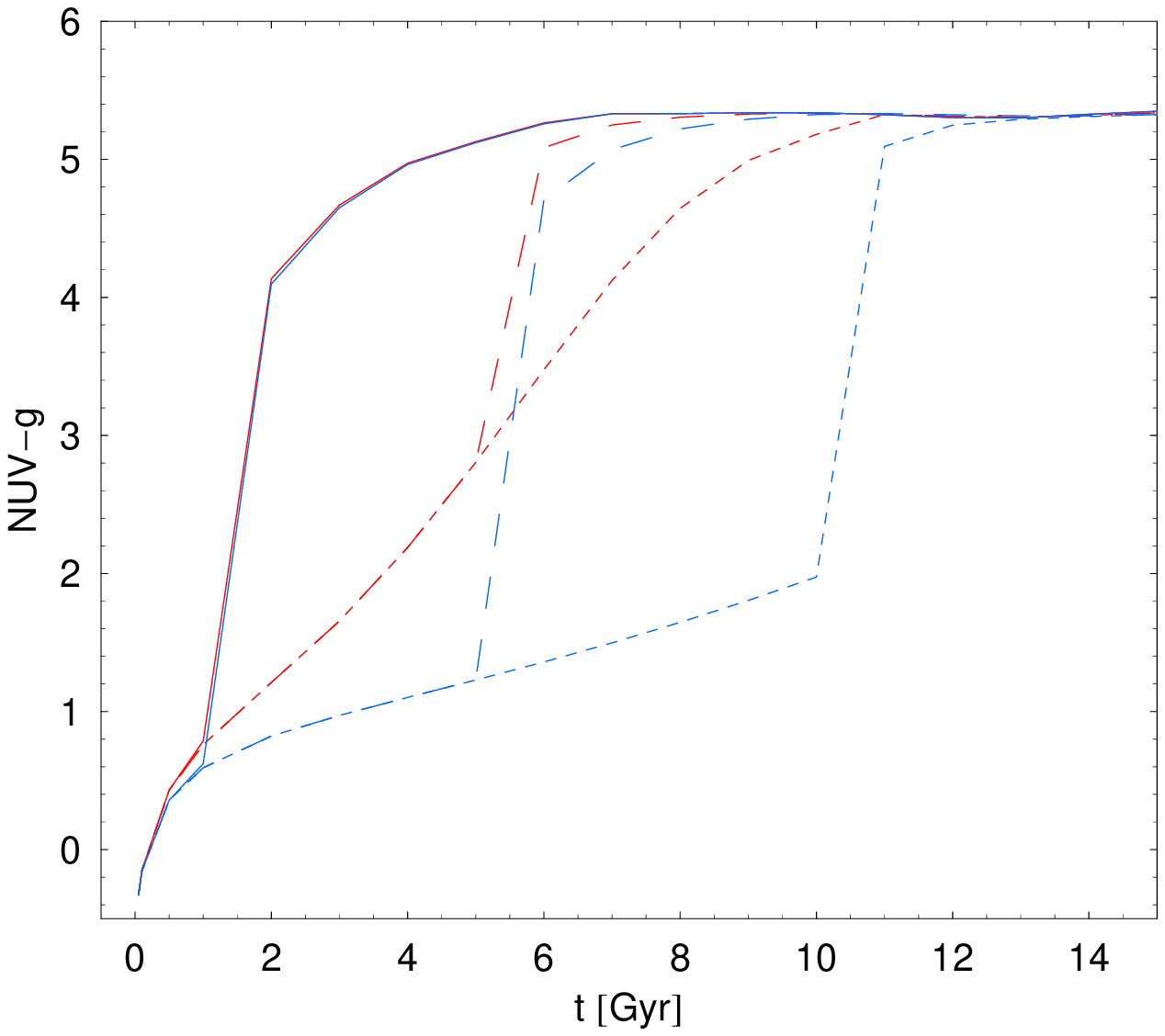,
width=0.32\textwidth}\hspace{0.2cm} \psfig{file=
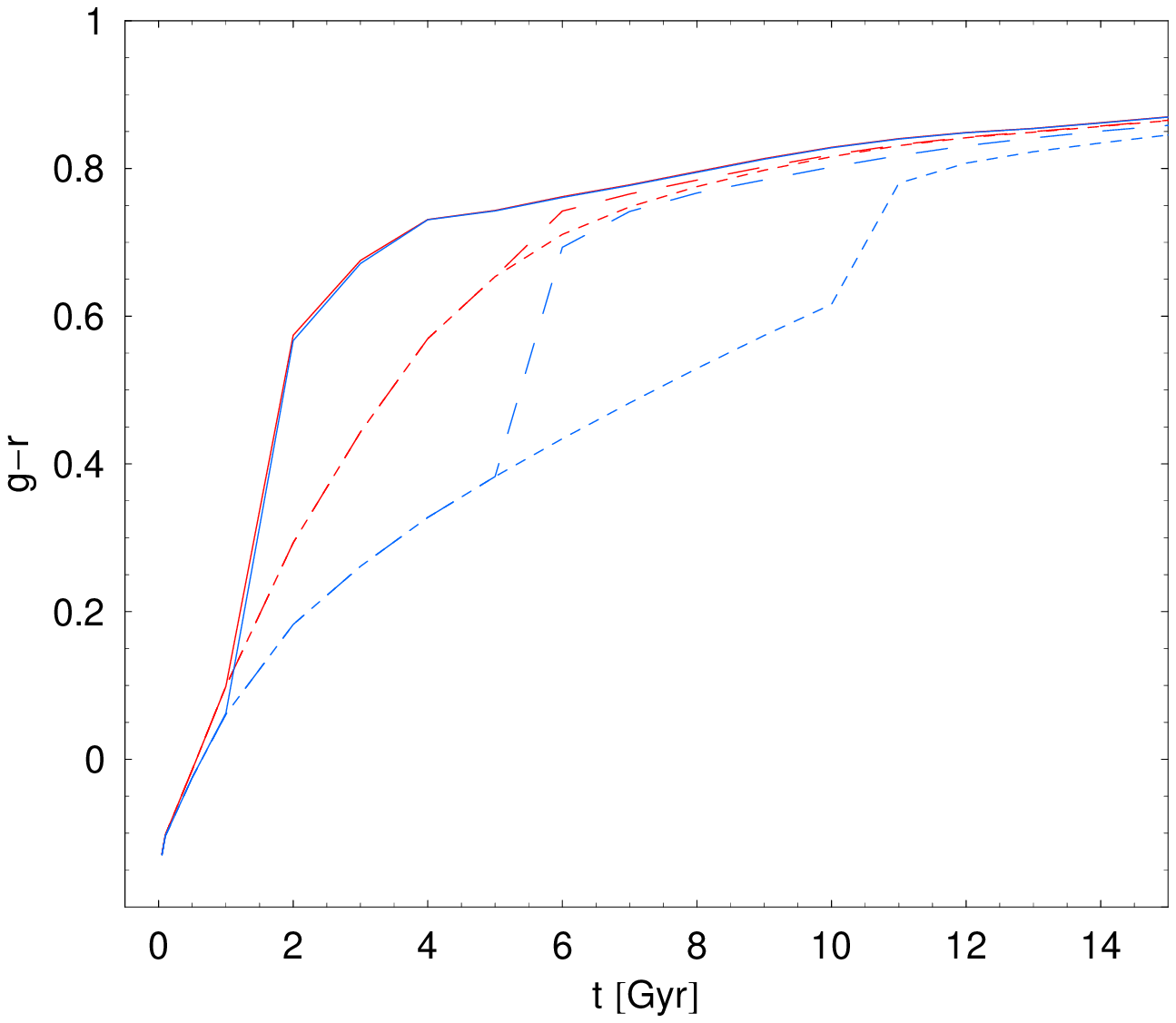, width=0.32\textwidth}\caption{Estimated
synthetic colours FUV-NUV, NUV-g and g-r (in AB system) as a
function of galaxy age (in units of $\rm Gyr$) for $Z=\Zsun$ and
for $\tau = 1 \, \rm Gyr$ (red lines) and $\tau = 3 \, \rm Gyr$
(blue lines). Continuous, long-dashed and short-dashed lines
correspond to $\tA = 1,\, 5,\, 10 \, \rm Gyr$. }
\label{fig:col_vs_t_2tau}
\end{figure*}

\begin{figure*}
\psfig{file= 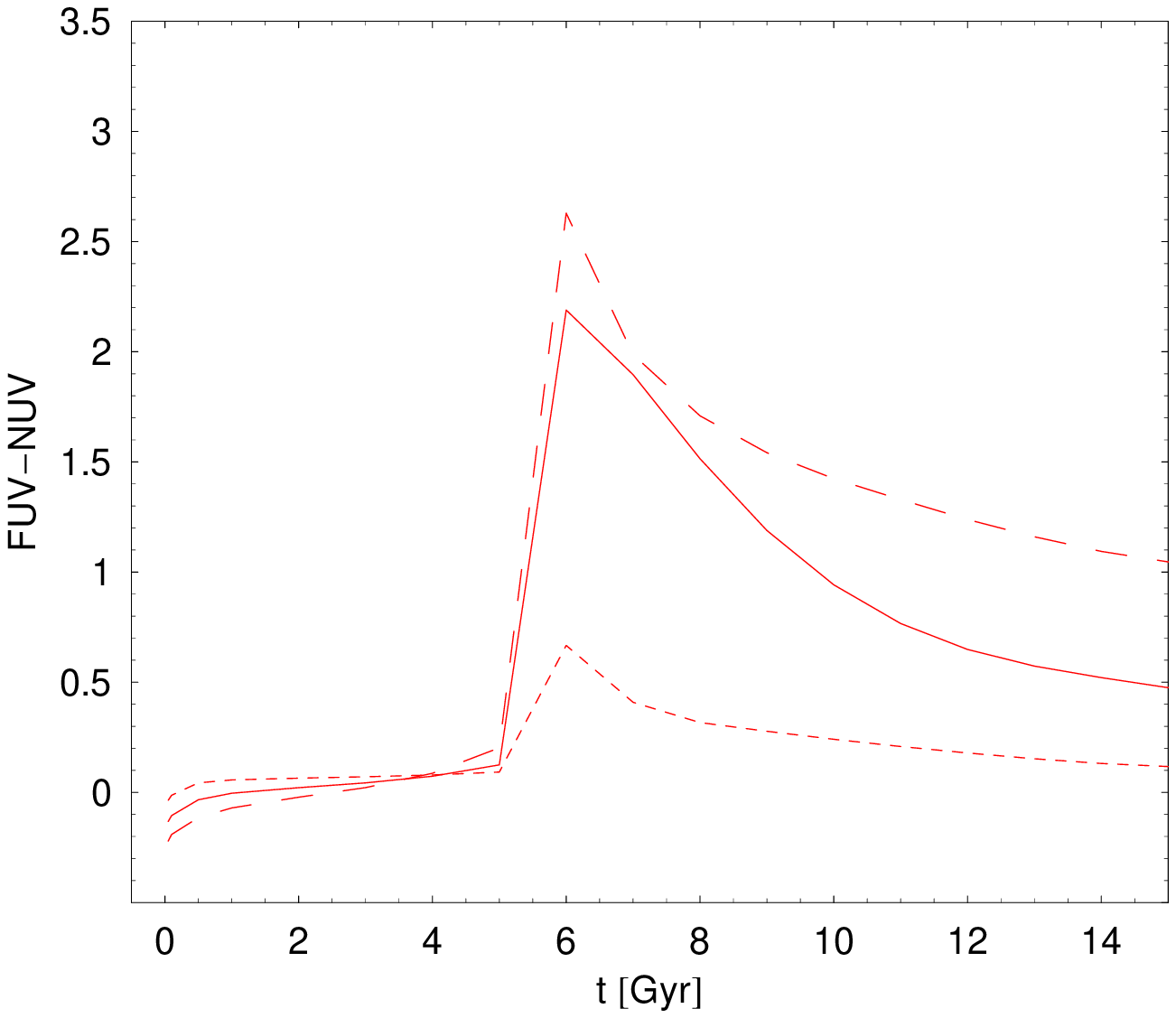,
width=0.32\textwidth}\hspace{0.2cm}\psfig{file= 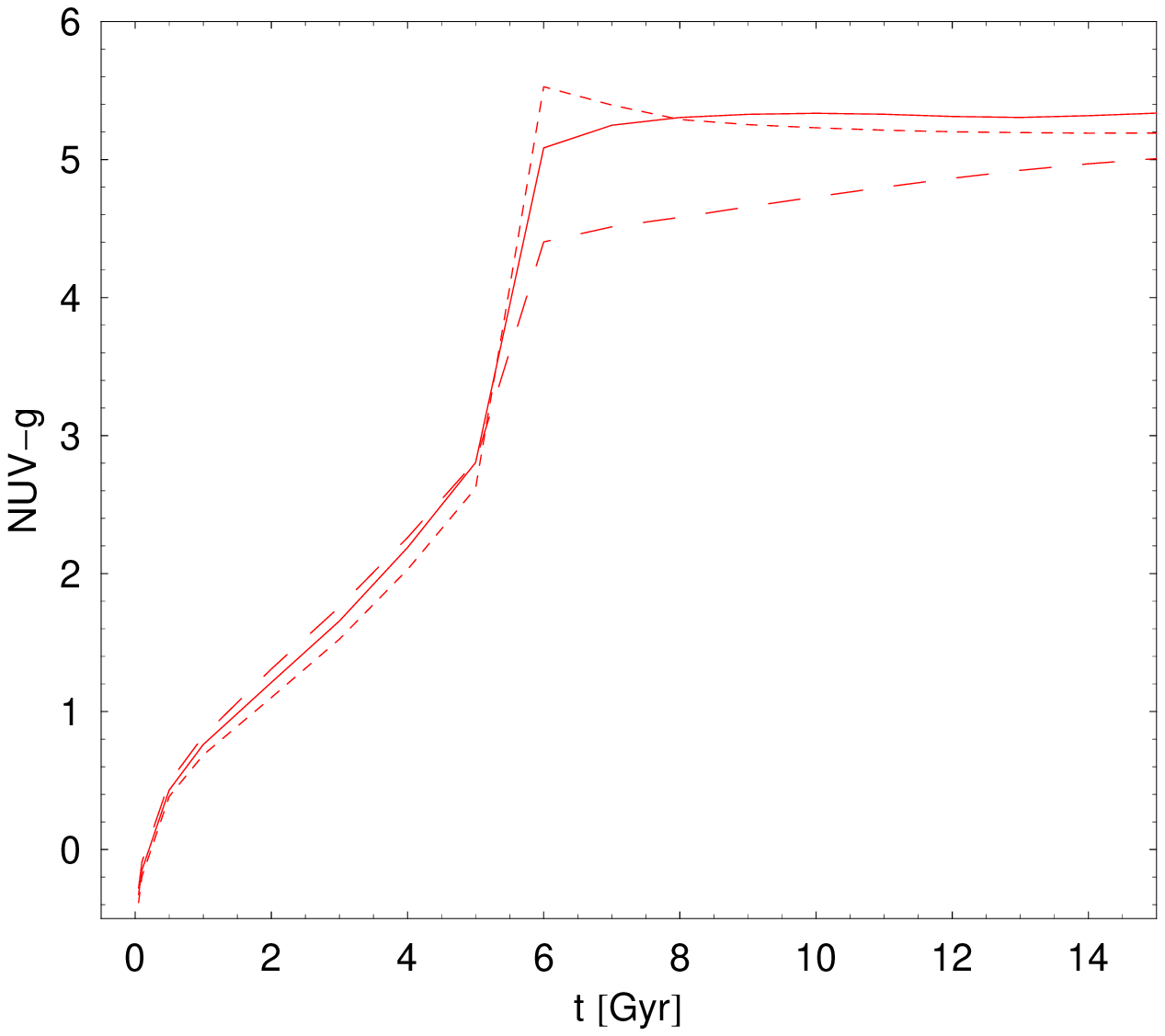,
width=0.32\textwidth}\hspace{0.2cm} \psfig{file=
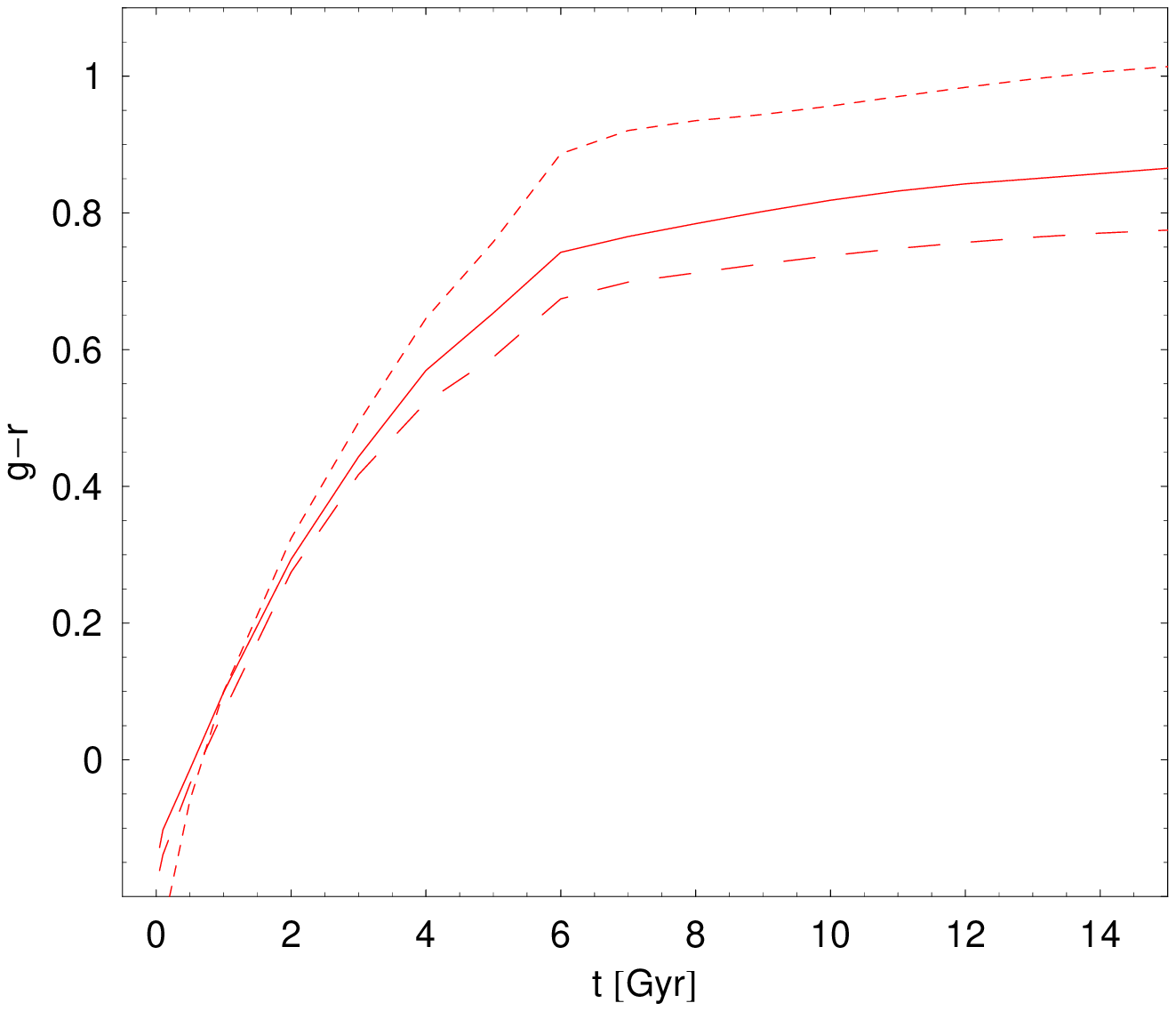, width=0.32\textwidth}\caption{Estimated
synthetic colours FUV-NUV, NUV-g and g-r (in AB system) as a
function of galaxy age (in units of $\rm Gyr$) for $\tau = 1 \,
\rm Gyr$ and $\tA = 5 \, \rm Gyr$. Long-dashed, continuous and
short-dashed lines correspond respectively to $Z=0.008, \, 0.02,
\, 0.05$.} \label{fig:col_vs_t_3Z}
\end{figure*}

\subsection{Time-scales}

Although we discuss the output of only one single simulation, the
general model of jet propagation in the ISM which is also probed
by our simulation allows us to determine one of the most relevant
parameters of the star feedback model  that we will develop in the
next paragraphs: the typical time-scale for suppression of stellar
formation.

In the self-similar expansion model of \cite{1997MNRAS.286..215K},
the cocoon is supposed to propagate into an unperturbed ISM which
is well described by a power-law profile: $\rho(r) =
\rho_{0}(r/a_{0})^{-\beta}$, and the typical scale length of the
jet varies with time according to: \be L_{j} =
c_{1}a_{0}\left(\frac{t}{t_{0}}\right)^{3/5 - \beta} \label{ts:1}
\ee where (\citealt{1997MNRAS.286..215K}, Eq. 5): \be t_{0} =
1.186 \times
10^{6}\left(\frac{a_{0}^{5}\,\rho_{0}}{P_{j,45}}\right)^{1/3}\,
{\rm yrs.} \label{ts:2} \ee
and $P_{j,45}$ is the jet's mechanical power in units of $10^{45} \, \rm erg \, s^{-1}$.\\
We have assumed that the number density of cold, star-forming
clouds is proportional to that of the diffuse gas, and thus can
be well approximated by the same power-law density profile outside
the core ($r>a_{0}$). There exists a threshold cloud number
density under which SF becomes negligible, and we will suppose
that this corresponds to a value of the gas density $\rho_{cr}
\approx 10^{-27} {\rm g \, cm}^{-3}$, which is reached at a
distance: $r_{cr} = a_{0}(\rho_{0}/\rho_{cr})^{1/\beta}$.
Inserting this into Eq.~(\ref{ts:1}), we find that the
characteristic timescale needed to reach this distance is given
by: \be t_{fb} =
t_{0}\left(\frac{\rho_{0}}{\rho_{cr}}\right)^{\frac{5}{\beta(3-\beta)}}c_{1}^{-\frac{5}{3-5\beta}}
\label{ts:3} \ee We plot in Fig.~\ref{fig:tscale} the dependence
of $t_{fb}$ on jet's power $P_{j}$. As we can see, typical values
for the time-scale of feedback are in the range $10^{6}-10^{8}$
yrs, for reasonable values of the input parameters. The values of
$P_{j}$ that we choose
are representative of the observed mass range of BHs at the
centers of typical galaxies, according to
the empirical relation between $M_{bh}$ and $P_{j}$ found by
\citet{2006ApJ...637..669L}. Thus, the quenching time-scale for
our simulation, of the order of a few Myrs, will increase of
about two orders of magnitudes for less powerful jets.

\begin{figure*}
\psfig{file= 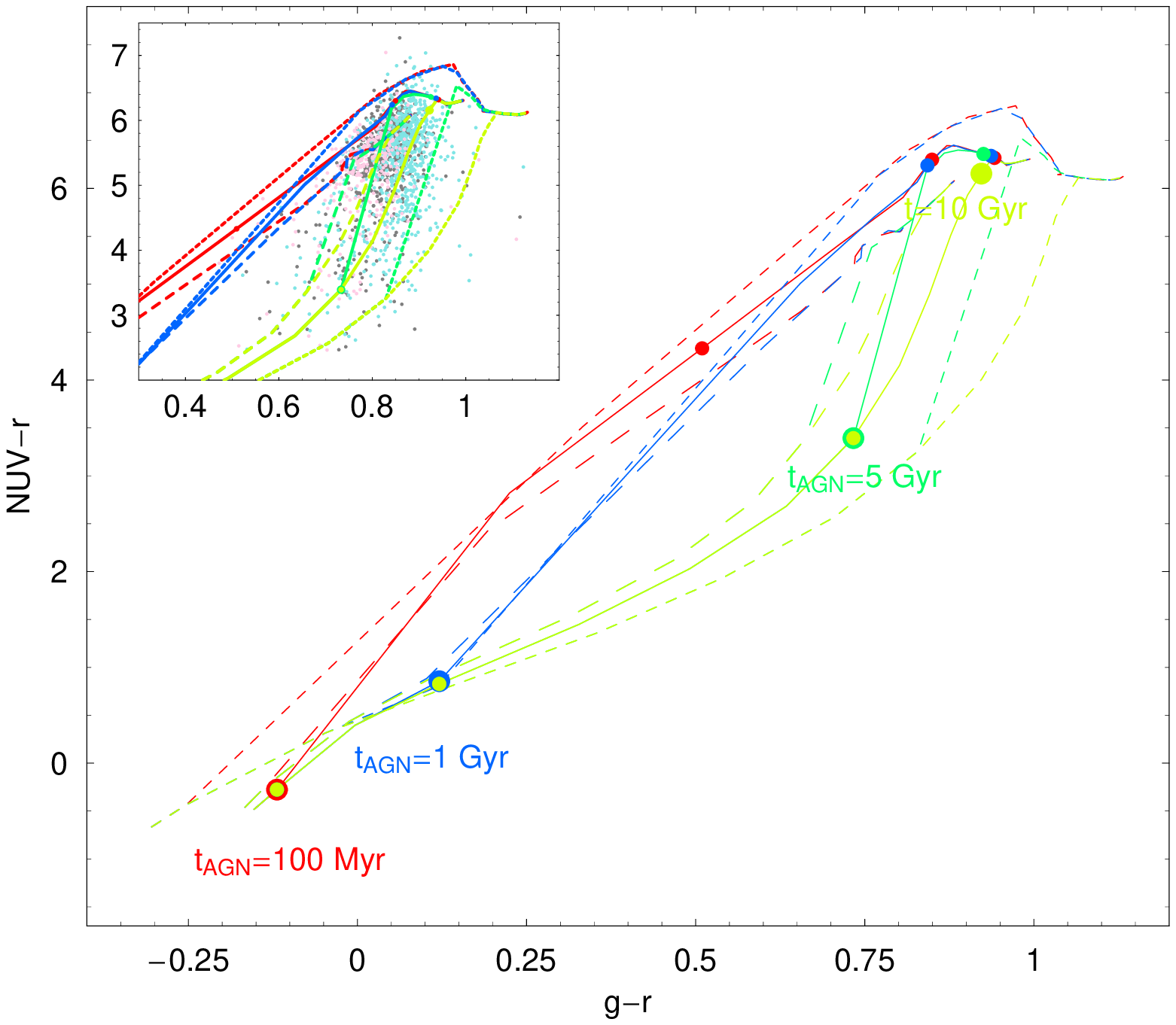, width=0.6\textwidth} \psfig{file=
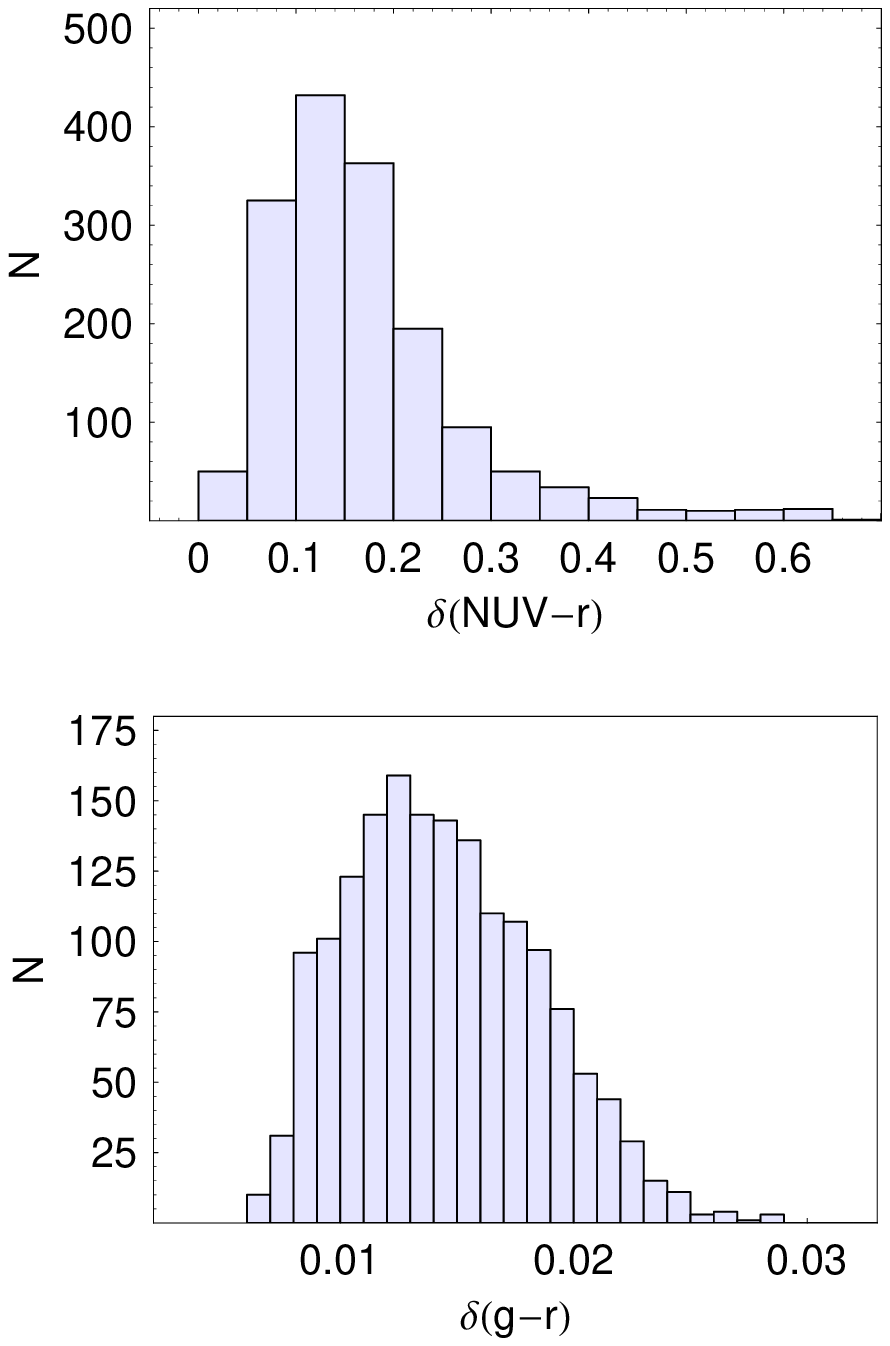, width=0.35\textwidth} \caption{UV/optical
colour-colour diagram (NUV-r vs g-r) from synthetic spectra
redshifted to median redshift of the sample $z_{med}=0.057$
assuming $\tau = 1 \, \rm Gyr$ and metallicities $Z=0.008$
(long-dashed line), $Z=0.02$ (continuous line) and $Z=0.05$
(short-dashed line). Red, blue, green and yellow lines correspond,
respectively, to $\tA =0.1, \, 1, \, 5, \, 10 \, \rm Gyr$.
The large points set the values of \tA\ on each synthetic track,
while the small ones indicate \tg\ ($=0.1, \, 1, \, 5, \, 10 \,
\rm Gyr$), which on different tracks correspond to different set
of colours after the quenching. In the inset panel we superimpose
 the sample galaxies to theoretical tracks, with violet, gray and
cyan points indicating galaxies in the redshift bins 0--0.04,
0.04--0.06 and 0.06--0.08. Instead of error bars, we show in the
right panels the distributions of uncertainties on galaxy colours
$g-r$ and $NUV-r$.} \label{fig:col_vs_col}
\end{figure*}

\section{Synthetic colours}\label{sec:colours}

The simulation discussed in this paper describes the impact
that the jet emitted by AGN has on quenching SF. However, we would
like to transpose this modification of the SF into a change of
some observable quantities. Colors and absorption/emission lines
are the most important observables given by the observations: here
we will concentrate on the former ones. In order to reproduce
galaxy colours, we use the synthetic models of \citet[hereafter
BC03]{BC03}, that allows us to encompass a wide range of
metallicities, starting from galaxy ages \tg\ of $10^{5} \, \rm
yrs$, and gives a full coverage in wavelength from $91 \, \rm
{\AA}$ to $160 \rm \, \mu m$. We start from a single initial
burst model, consisting of a single population with a Chabrier
IMF\footnote{Galaxy colours are unchanged if we use a Salpeter
IMF, since these two IMFs differently describe the distribution of
low mass stars that contribute little to the light distribution,
while strongly affecting the total stellar mass
(\citealt{Tortora09}).} and three values of the metallicities:
$Z=0.008, \, 0.02$ i.e. ($Z = \Zsun$), $0.05$ and convolve it with
a SFR law.

We suppose that without interaction of the AGN with the ISM, SF
within the clouds evolves following an exponential SF law
$SFR \propto e^{-t/\tau}$, where $\tau$ is a characteristic
timescale corresponding to the time when SF is reduced of
$e^{-1}$.. Then at $t=\tA$, the AGN interaction begins to work,
shocking the ISM and inhibiting SF. This configuration can be
realized by combining the unperturbed exponential SFR for $t \leq
\tA$ and SFR from our simulation for $t > \tA$. Due to the short
time-scale involved in the AGN effect on SFR, positive feedback is
not observable in the derived colours, while the negative feedback
is directly translated into a strong truncation of SFR. Note that
this model is an approximation, since less powerful AGNs are
observed at low redshifts and the decrease of SF can be shallower,
as discussed in the previous section (see, also, the
parametric model of SF quenching discussed in \cite{Martin07}).
If $\tau \to \infty$, the SFR becomes a burst with finite length
\tA\ (i.e. a SF constant up to \tA\ with a null value for $t >
\tA$).

We calculate colours by convolving the filter responses of $\rm
u,\, g,\, r,\, i$ and z SDSS and the ultraviolet FUV and NUV GALEX
bands (\citealt{Martin05}, \citealt{Yi05}, \citealt{Kaviraj07})
with synthetic spectra. In Figs. \ref{fig:col_vs_t},
\ref{fig:col_vs_t_2tau} and \ref{fig:col_vs_t_3Z} we show the
change with time of colours $FUV-NUV$, $NUV-g$ and $g-r$ and the
effect of quenching. In particular, in Fig. \ref{fig:col_vs_t} we
compare the colours of our inhibited SF (with $\tA= 1$ and $5 \,
\rm Gyr $) and two simple models using an exponential SF with
$\tau=0$ and $\tau = 1 \, \rm Gyr$. SF inhibition essentially
reddens the colours starting at t$>$\tA. Ultraviolet is much
sensitive to SF inhibition (see the middle panel in Fig.
\ref{fig:col_vs_t}), with variations of $1-3$ magnitudes
in $NUV-g$; less sensitive are the visible colours, with variations of
$\sim 0.2-0.4$ in $g-r$. The colour FUV-NUV increases after the
action of the AGN up to $\sim 1-4$, depending on \tA, but after
this event it decreases reaching a value $\lsim 1$ for large \tg.
Therefore, the latter colour is degenerate, because one is not able to
univocally determine the galaxy age from colour (each colour
corresponds to two possible values of \tg).

A single burst model ($\tau = 0$) is similar to our models with
inhibited SF for $t > 2-4 \, \rm Gyr$, but predicts larger values
for $t < 2-4 \, \rm Gyr$. While $g-r$ colours for the different
models shown in Fig. \ref{fig:col_vs_t} are similar since $t \sim
6-8 \, \rm Gyr$, the other colours are comparable at larger
ages $t \sim 8-10 \,$ Gyr, being more sensitive to SF. Not
shown in the figure is the case $\tau \to \infty$, characterized
for fixed values of \tA\ and \tg\  by bluer colours.

These considerations are obviously dependent on the choice of
$\tau$ and $Z$. In Fig. \ref{fig:col_vs_t_2tau}, we compare the
results for $\tau = 1$ and $3  \, \rm Gyr$. At fixed age, a more
protracted unperturbed SF predicts smaller $NUV-g$ and $g-r$ and a
larger $FUV-NUV$. Metallicity has the opposite effect: $FUV-NUV$
decreases if we increase metallicity, while g-r increases;
finally, a more complex behaviour is observed for $NUV-g$, as can
be seen by inspecting Fig. \ref{fig:col_vs_t_3Z}. These complex
dependencies on metallicity are linked to the details of stellar
population prescriptions.

We can outline a picture where the AGN effect has a main role in
the evolution of brighter elliptical galaxies. In particular,
inhibition of SF by AGN transforms a galaxy with a protracted
SF (in our case an unperturbed SF obtained using an exponential SF
law with a specific $\tau$) into a more quiescent galaxy with SF
stopped at a time approximately equal to the epoch of jet injection. The extreme case of this model is represented by a single
burst model, where only at $t=0$ is the SF observable and most of
the SF has been completed by this epoch.

\section{Epoch of quenching event}\label{sec:Observations}

We will now attempt to fit synthetic models to a large sample of
local ETGs, in order to obtain information about the epoch of
quenching event and RSF. The galaxy sample is presented in Sect.
\ref{sec:sample}, while the spectral fit procedure and the first
results are described in Sect. \ref{sec:fit}. We go into more
detail in Sects. \ref{sec:qe} and \ref{sec:RSF}, where quenching
and the properties of the recemtly formed stellar populations are
discussed. Finally, few
comments on galaxy evolution are addressed in Sect.
\ref{sec:evolution}.

\subsection{Galaxy sample}\label{sec:sample}

We use a sample of ETGs extracted from the SDSS, using a selection
procedure described in \cite{Kaviraj07}. The initial selection is
made using the {\it fracDev} parameter in SDSS, that attributes a weight
 to the best composite (deVaucouleur's +
exponential) fit to the galaxy image in a particular band. The
criterion {\it fracDev} $> 0.95$ has been proven to be extremely
robust, allowing one to pick up $\sim 90 \%$ of ETGs in a typical
sample of SDSS galaxies. Furthermore, a visual inspection of SDSS
images is needed to refine the selection: the ability to classify
galaxies obviously depends on redshift and apparent magnitude of
the observed galaxies. In order to construct a magnitude-limited
sample, we restrict ourselves to r-band magnitude $< 16.8$ and
redshift $z<0.08$. Finally,  cross-matching with ultraviolet
GALEX data produces the final sample, with measured magnitudes in
SDSS bands u, g, r, i, z and GALEX  FUV and NUV (see
\cite{Kaviraj07} for further details). In this way, we have a wide
coverage of galaxy spectra, since in the optimal cases,
magnitudes in 7 bands are observed, ranging from $\lambda \sim
1500 \, \rm \AA$ up to $\lambda \sim 9000 \, \rm \AA$. SDSS
magnitudes have typical uncertainties of $\sim 0.01$, while GALEX
data are more uncertain with mean errors of $\sim 0.25$ and $\sim
0.15$, respectively for FUV and NUV magnitudes.

\begin{figure}
\psfig{file= 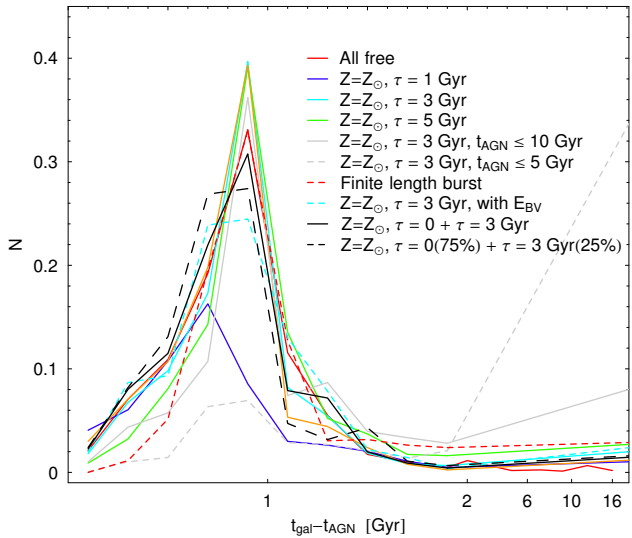, width=0.47\textwidth}
\caption{Distribution of recovered best fit values of $\tg - \tA$
for different spectra sub-libraries extracted from our initial
library. On the y-axis, we plot the fraction of galaxies
(normalized to dimension of the sample) within bins of $\tg -
\tA$. In the range $\tg - \tA \in [0-2] \, \rm Gyr$ the scale is
linear, while for $\tg - \tA > 2 \, \rm Gyr$ we use an arbitrary
logarithmic scale. In the linear range we group galaxies in bins
of size $0.2 \, \rm Gyr$, while larger bins are used outside this
range. In particular, we show the results leaving all parameters
free to change, setting $Z = \Zsun$ and $\tau = 1$, $3$ or $5 \,
\rm Gyr$, for $\tau = 3 \, \rm Gyr$ we analyze two cases with
constraints on \tA\ and then include internal extinction in the
fit. We analyze the case of a finite length burst ($\tau =
\infty$) and a combination of two coeval stellar populations with
$\tau=0$ and $\tau = 3 \, \rm Gyr$ with a free mass ratio or a
fixed one. See labels in the plot and discussion in the text for
further details.} \label{fig:Distr_various_fits}
\end{figure}

\begin{figure}
\psfig{file= 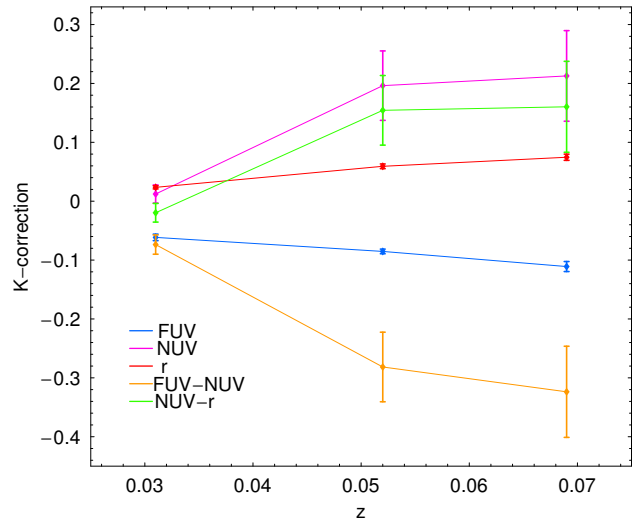, width=0.47\textwidth} \caption{
Median K-corrections derived by fitted synthetic spectra for
bands FUV, NUV, r and colour FUV-NUV and NUV-r. The bars are the
median scatters. See legend for details on the colour code.}
\label{fig:K_corr}
\end{figure}

In Fig. \ref{fig:col_vs_col} we show a colour-colour ($NUV-r$ vs
$g-r$) diagram from our synthetic tracks. In the inner plot we
present our sample, superimposing the galaxies, colour coded according
to redshift
bins. This diagram gives information about both \tg\ and \tA\,
selecting specific values for two colours and the best parameters
of galaxies. Different values of galaxy parameters predict galaxy
colours that populate different regions of the diagram.

The sample of \cite{Kaviraj07} is suitable for our analysis,
since AGN spectral features are observed in spectra of many
galaxies. Type-I AGN are automatically removed by using the SDSS
spectral classification algorithm, while we are interested in
galaxies that host a Type-II AGN. To distinguish between normal
star forming galaxies and galaxies hosting an AGN, it is usual to
analyze a few strong emission lines, e.g. the emission line ratios
$[OIII/H\beta]$ and $[NII/H\alpha]$ (\citealt{BPT81},
\citealt{K2003}). Galaxies with both these indicators measured represent
$\sim 65\%$ of the galaxy sample, and $\sim 86\%$ of them have
spectral features consistent with those of LINER, Seyfert or
transition objects.

Emission from Type-II AGN does not affect the stellar continuum of
host galaxies. In fact, for luminous Type-II AGNs the maximum
contamination in flux amounts to few percents in visible bands and
to less than $15\%$ (translating to 0.15 mags) in UV bands
(\citealt{K2003}, \citealt{Salim07}). In addition, galaxies
hosting AGN are systematically redder in the UV colours than their
counterparts which do not have AGNs, further on suggesting that
continuum emission from AGN leaves unaffected galaxy spectra.

\subsection{Spectral fit}\label{sec:fit}

We build a library of synthetic spectra, using our SFR
prescription with galaxy age \tg, $\tau$, \tA\ and $Z$ as free
parameters. We divide the sample into  3 redshift bins 0--0.04,
0.04--0.06 and 0.06--0.08 and we move our spectra to the median
redshift of each bin (respectively of 0.031, 0.052, 0.069).
Synthetic magnitudes and colours are obtained convolving these
redshifted spectra with the filter responses. Finally, the
synthetic colours are fitted to the observed ones ($FUV-NUV$,
$NUV-g$, $u-g$, $g-r$, $g-i$ and $g-z$), by a maximum likelihood
method, which allows us to estimate the best values for the free
galaxy parameters. In this way we do not need to have
K-corrections, leaving the fit safe from possible uncertainties
introduced by these corrections, and the simplistic division of
our sample into only 3 redshift bins will not affect our
estimates. Later, we will discuss the K-correction that we derive
from the fitting procedure and we will use to obtain rest-frame
magnitudes.

\begin{figure*}
\psfig{file= 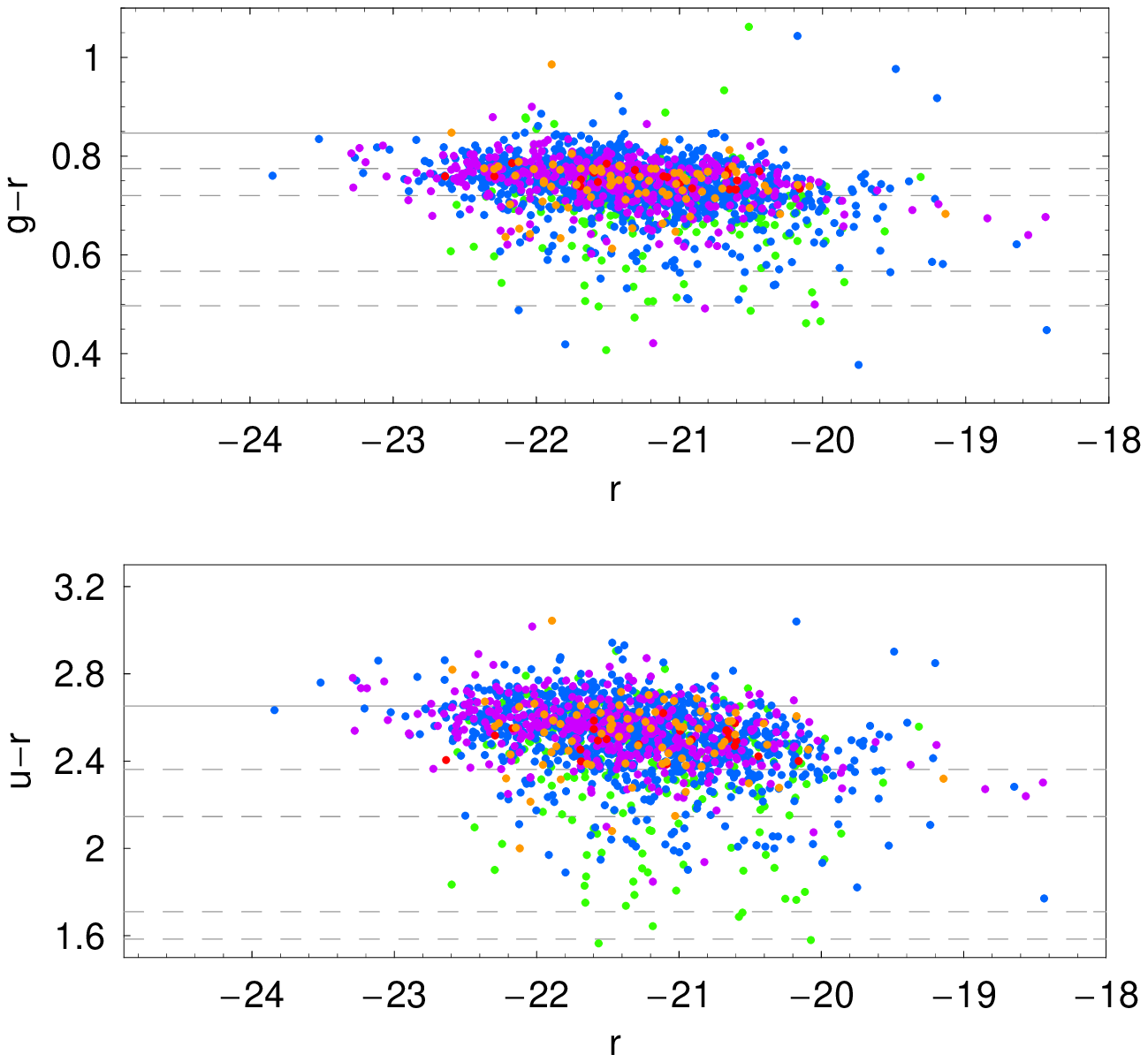, width=0.50\textwidth}\psfig{file=
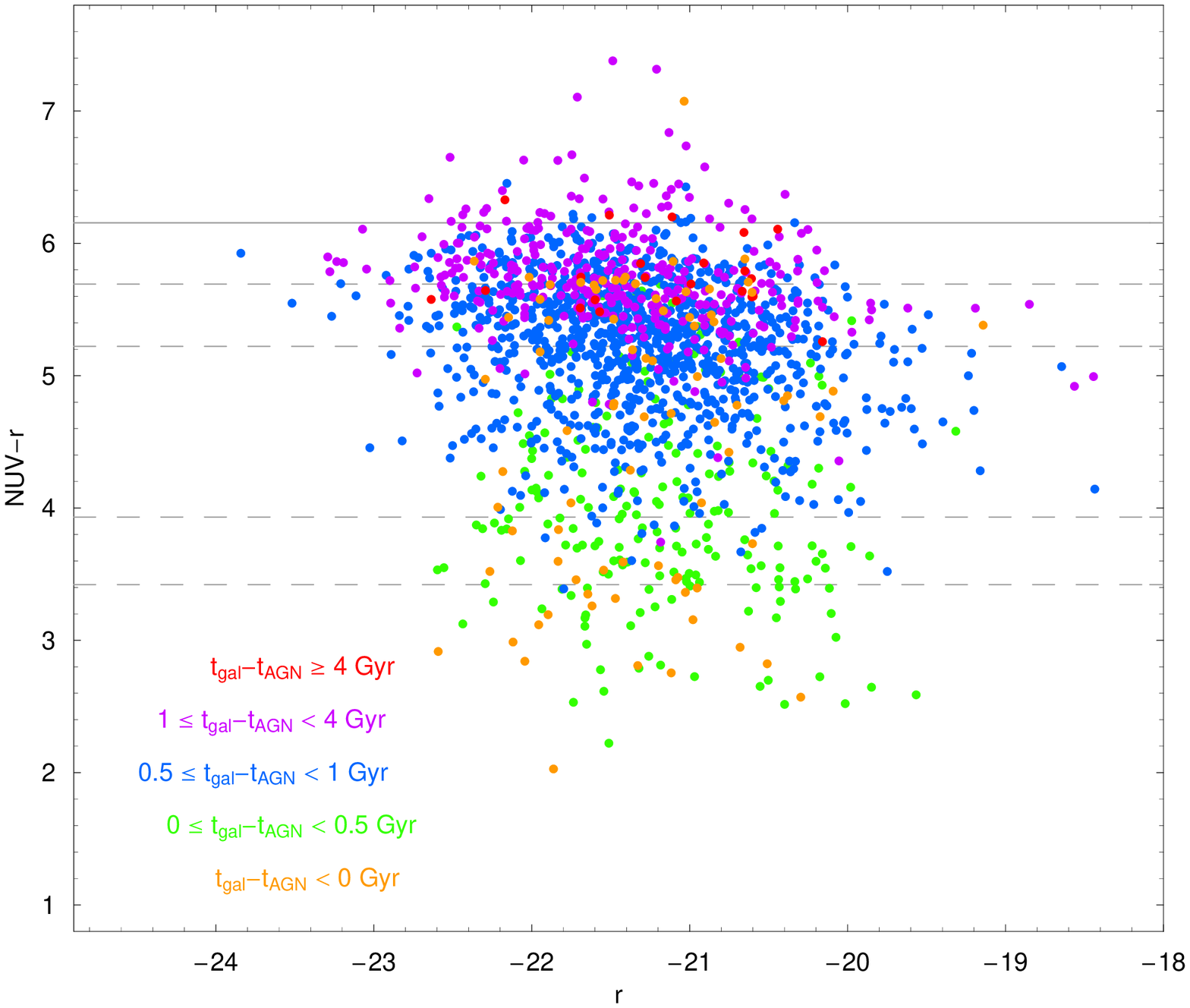, width=0.49\textwidth}
\caption{Colour-magnitude diagrams of our sample, colour-coded
following the classification shown in the legend: $\tg - \tA <0$
(orange), $0\leq \tg - \tA < 0.5$ (green), $0.5 \leq \tg - \tA <
1$ (blue), $ 1 \leq \tg - \tA < 4$ (violet), $4 \leq \tg - \tA$
(red). We show $z=0$ colours predicted using a $z_{f}=3$ single
burst population for $Z=\Zsun$ as a continuous gray line and
various $Z < \Zsun$ (i.e. $Z=0.0001,\, 0.0004, \, 0.004, \,
0.008$) as dashed gray lines. {\it Left Panels}. We plot g-r vs r
and u-r vs r diagrams. {\it Right Panel}. We plot NUV-r vs r
magnitude.} \label{fig:NUVr_vs_r_data}
\end{figure*}

We restrict synthetic library to spectra with \tg\ free to change
(with a tiny step) up to $17 \, \rm Gyr$, $\tA \in (0, 15)\, \rm
Gyr$, $Z \in (0.008, \,0.02, \,0.05)$ and 3 reference values for
the SFR scale $\tau = 1,\,3, \,5 \, \rm Gyr$. For each value of
\tg\ we have more than 150 spectral models. This library is wide
enough to reproduce spectral features of ETGs
(\citealt{Panter08}). The range of metallicities used here has
been shown to be representative of luminous ETGs with $M_{B} \lsim
-19$ (\citealt{Romeo+08}, \citealt{Tortora09}). In particular, the
fit of spectra with an unperturbed exponential SF and variable $Z$
to local ETGs gives on average values of $\tau \lsim 1$ and $Z
\geq Z_{\odot}$ (\citealt{Tortora09}). Such an unperturbed
exponential SF with a low time-scale $\tau$ can disguise a more
complex SF evolution, which we have obtained from the simulation
and want to probe against observations.

Before proceeding to model the properties of the observed sample of galaxies,
we briefly discuss the possible systematics which the fitting
procedure can generate in the best fitted parameters, by
performing a set of Montecarlo simulations on synthetic colours.
We have extracted a large sample of simulated spectra from our SED
library with random \tg, \tA, $\tau$ and $Z$. Then, we applied our
fitting procedure and compared the recovered best fit parameters
against the intrinsic ones. While \tg, \tA\footnote{Note that the
input value for \tA\ is correctly recovered if $\tg - \tA \geq 0$,
while for $\tg - \tA < 0$ it is not possible to constrain its
value.} and $Z$ are recovered quite well, $\tau$ is poorly
constrained. To reduce the unavoidable degeneracies in the fitting
procedure, it would be acceptable to set $\tau$ to a fixed value,
however, in the following we will discuss both the cases with
$\tau$ variable and constant. In addition, to further reduce the
{\it noise} in our final results we also analyze the effect of
adopting constant metallicity.

Coming back to the observed sample of galaxies, we use
different synthetic libraries to fit the observed colours,
extracting spectra from our initial and extended set of synthetic
models. We show the results of this analysis in Fig.
\ref{fig:Distr_various_fits}, where we plot the distribution of
recovered values of $\tg - \tA$. If on one hand the change of
spectral library can modify the estimates of single values of the
parameters, on the other hand it leaves the main clump of the
distribution of differences $\tg\ - \tA$ unaffected. Due to the
correlations among parameters and constraints imposed on some of
them, the sample distributions of recovered values for \tg, \tA\
and Z can change, but median values for $\tg\ - \tA$ ($\sim 0.8 -
0.9\, \rm Gyr$ with a median scatter of $\sim 0.1-0.2 \, \rm Gyr$)
are almost unchanged, with a scatter among the different estimates
consistent with 0. If we impose strong constraints on \tA\ (e.g.,
$\tA\ \leq 5 \, \rm Gyr$), we observe a departure from the mean
trend obtained using other libraries, with a peak at very large
values of $\tg - \tA$ (corresponding to quenching at high
redshift), but this constraint does not seem to be motivated since
it decreases the quality of fit. In addition, note that for $\tau
= 1 \, \rm Gyr$ the distribution peak is slightly lower and less
prominent, mainly due to the larger number of galaxies estimated
to have unperturbed SF (i.e., with $\tg < \tA$). To perform
a more complete analysis, we also show the results for a combined
spectral model obtained by summing up two coeval stellar
populations: a single burst model to a quenched SF assuming a
variable or a fixed mass ratio of the two populations. This model
takes into account the coexistence of stellar populations
having different properties: a part of stars formed in a single
initial burst, in addition to this component, there is gas which
cools to form continuously stars and is affected by AGN.
What we see in Fig. \ref{fig:Distr_various_fits} is that these
combined models we have discussed give the same results as using
the single component ones. Our estimates of $\tg - \tA$ are
robust, since the bulk of their distribution depends little on the
duration of the unperturbed SF and $Z$. Despite this result, as we
will see later, the past SF history of galaxies is dependent on
the choice of these models. Finally, two shortcomings have to be
discussed. If considering more protracted unperturbed SF we will
obtain a larger number of galaxies with higher $\tg - \tA$ values
than using SFs with lower $\tau$ (see Fig.
\ref{fig:Distr_various_fits}). In addition, a similar effect is
given if quenching is not almost instantaneous as in our
simulation, also in this case we would obtain larger values of
$\tg - \tA$. However, our fast quenching is a good approximation
to describe AGN effect for a wide sample of galaxies, but further
analysis on these aspects waits to be done in future.

In order to avoid confusion, we will use in the following the
results obtained in the most general case where $Z$, $\tau$, \tg\
and \tA\  are all unconstrained. Also, K-corrections deserve
particular attention. In fact, due to uncertainties in the UV
spectrum, it  can be difficult to quantify the amount of this
correction, acircumstance which can introduce  systematics. In our
range of redshifts, we find typical NUV corrections of $0-0.3$, in
good agreement with those found in \cite{Kaviraj07} using the same
galaxy catalog. In Fig. \ref{fig:K_corr}, we show also results for
K-corrections of FUV, r and colours FUV-NUV and NUV-r. We obtain a
sharp rise in NUV correction: this is negligible in the lowest
redshift bin, while its median value becomes $\sim 0.25$ at higher
z. The median scatter in the estimated values of NUV and FUV
corrections is  consistent with typical uncertainties of observed
magnitudes in these bands. Note that \cite{Kaviraj07}, using a $9
\, \rm Gyr$ old single burst model and \cite{Rawle08}, using both
single burst and frosting models, find larger corrections of
$0.2-1.2$ for galaxies having $z \lsim \, 0.1$.

\subsection{Quenching event}\label{sec:qe}

In Fig. \ref{fig:NUVr_vs_r_data} we show the visible and UV
colour-magnitude diagrams for our sample. As already detailed in
\cite{Kaviraj07}, the UV diagram in the right panel of this figure
is largely different from similar colour diagrams for optical
bands (see left panels), since it shows a larger scatter. Fig.
\ref{fig:NUVr_vs_r_data} shows the power of $NUV-r$ colour in
selecting galaxies with different SFR histories. In particular,
the points in the figure are colour-coded according to the fitted
value of $\tg - \tA$: galaxies that experienced some quenching
phenomena in the past lie upper in this diagram, while those
having lower values of this quantity populate on average a region
with a flatter UV continuum and a larger scatter.

\begin{figure}
\psfig{file= 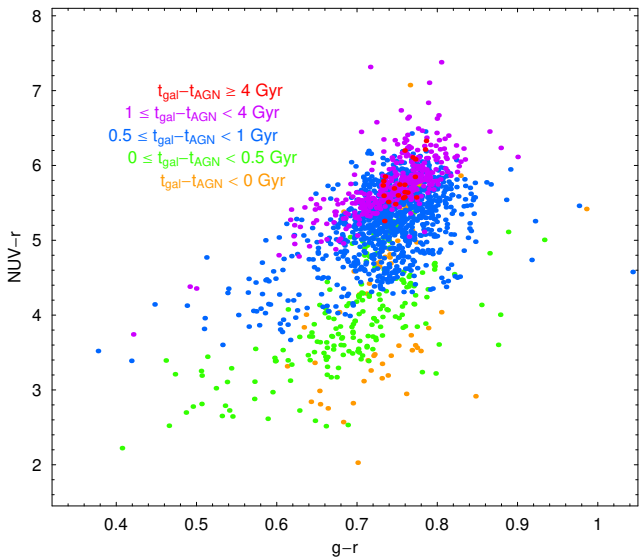, width=0.49\textwidth}
\caption{UV/optical colour-colour diagram. We plot NUV-r vs g-r of
galaxies in our sample colour-coded following the classification
shown in the legend.} \label{fig:NUVr_vs_gr_data}
\end{figure}

\begin{figure}
\psfig{file= 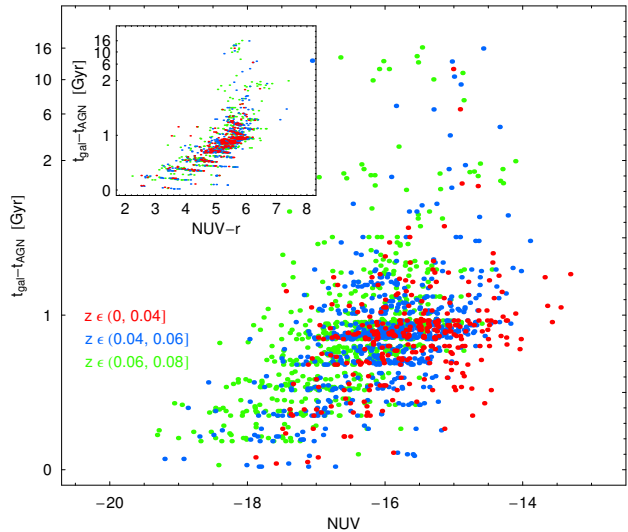, width=0.49\textwidth}
\caption{$\tg-\tA$ as a function of ultraviolet magnitude NUV,
while in the inserted panel is plotted $\tg-\tA$ as a function of
colour NUV-r. Points are colour coded for various redshift bins.
See Fig. \ref{fig:Distr_various_fits} for details on the
scale used in the y-axes.} \label{fig:tgtA_vs_NUVr_NUV}
\end{figure}

A rough comparison with the expectation of the monolithic scenario
has been made,  superimposing the predicted colours for single
burst populations with a more extended formation redshift starting
at $z_{f}=3$. For completeness, in addition to a model with
$Z=\Zsun$ we also show the predictions for various metallicities
$< \Zsun$ implemented within BC03 prescription. These predictions
are not very sensitive to the precise value of $z_{f}$ (at least
when considering relatively old stellar populations), and our
choice is consistent with optical analyses which estimate $z_{f}
\gsim \, 2$ (\citealt{Bower92}). We see that an almost solar
single burst model is only consistent with redder galaxies both
for g-r and NUV-r colours. Thus, while galaxies with low g-r and
NUV-r can be matched by models with low $Z$, galaxies in the u-r
vs r plot appear to depart from these predictions which
systematically lead to redder colours. However, as we pointed out
in a previous subsection, luminous ETGs are almost solar and very
low values of metallicities like those used in Fig.
\ref{fig:NUVr_vs_r_data} are not applicable to our galaxies
(\citealt{Panter08}, \citealt{Tortora09}). Moreover, semi-analytic
simulations predict that a single solar metallicity is a
reasonable hypothesis to describe ETGs (\citealt{NO06}) and
similar results for the brighter and more massive galaxies have
been obtained in N-body+hydrodynamical simulations
(\citealt{Romeo+08}). Thus, the comparison with a solar
metallicity single burst model indicates that only a few galaxies
($\lsim \, 6-7 \%$) are consistent (within the errors) with a
single and old stellar population; these galaxies are the redder
ones mainly with $\tg -\tA \gsim \, 1$. This result is consistent
with the $\lsim \, 1 \%$ of purely passive galaxies found in
\cite{Kaviraj08} analyzing a sample of galaxies at $z \gsim \,
0.5$, since in the local universe we observe a larger number of
quiescent and passive galaxies.

In Fig. \ref{fig:NUVr_vs_gr_data} we show NUV-r vs g-r, with
points colour-coded according to Fig. \ref{fig:NUVr_vs_r_data}. A
correlation is observed, but note that the excursion in NUV-r is
much larger than in g-r.

\begin{figure*}
\psfig{file= 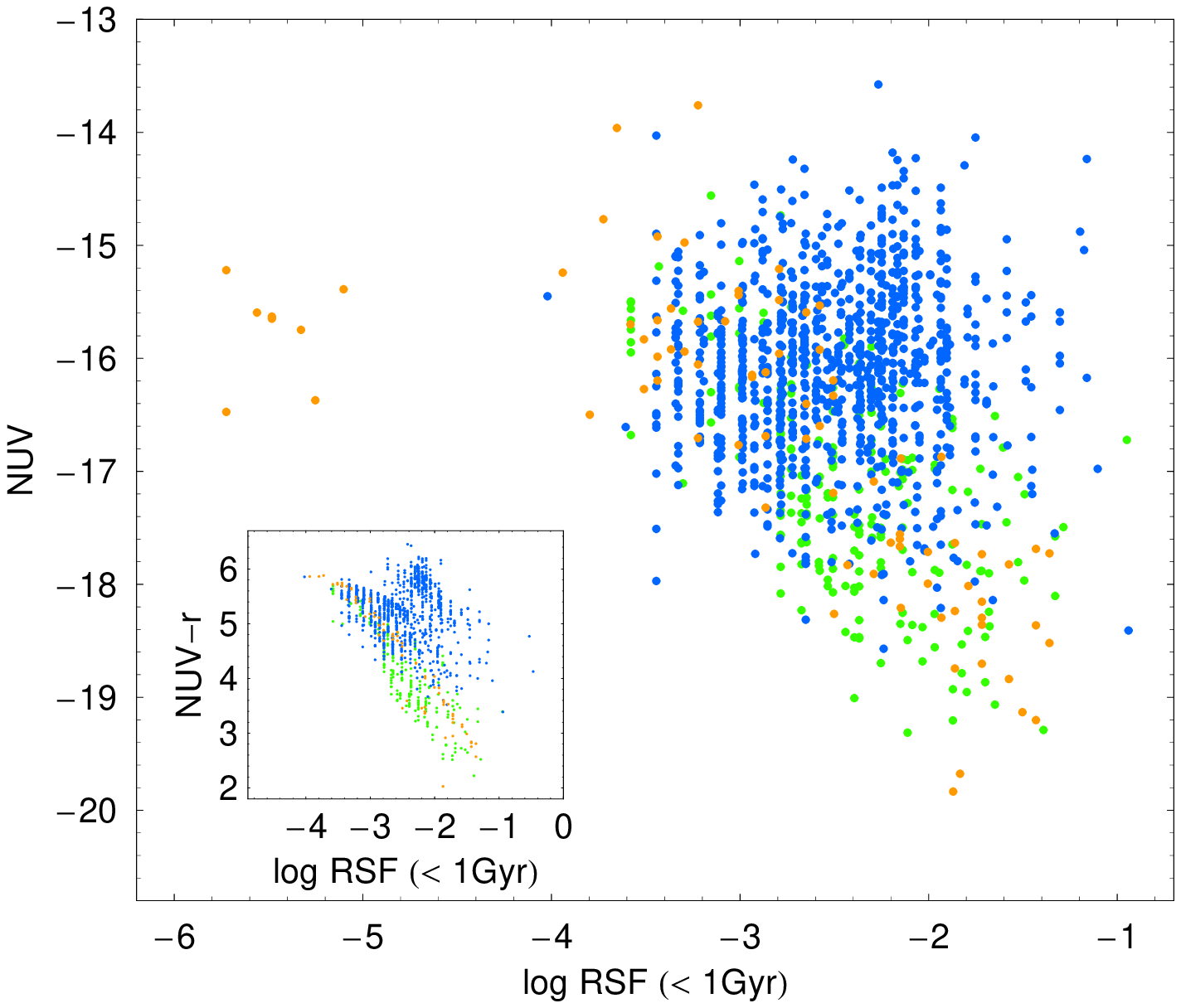, width=0.49\textwidth}\psfig{file=
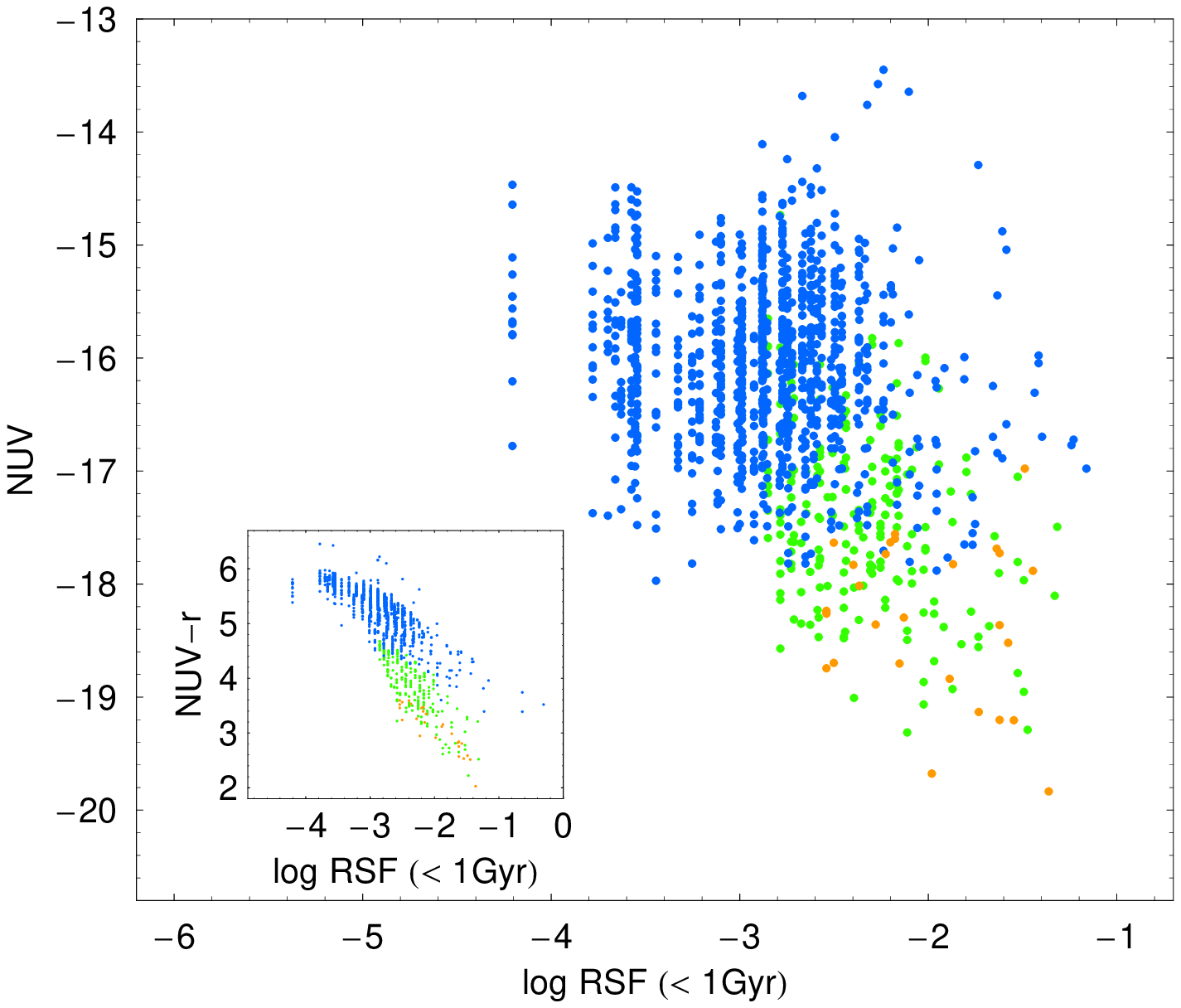, width=0.49\textwidth} \caption{NUV (and NUV-r)
vs. RSF for a model leaving free all parameters (left panel) and
one with $Z=\Zsun$ and $\tau=3 \, \rm Gyr$ (right panel). The
colour code is the same as in Figs. \ref{fig:NUVr_vs_r_data} and
\ref{fig:NUVr_vs_gr_data}.} \label{fig:RSF1}
\end{figure*}

\begin{figure*}
\psfig{file= 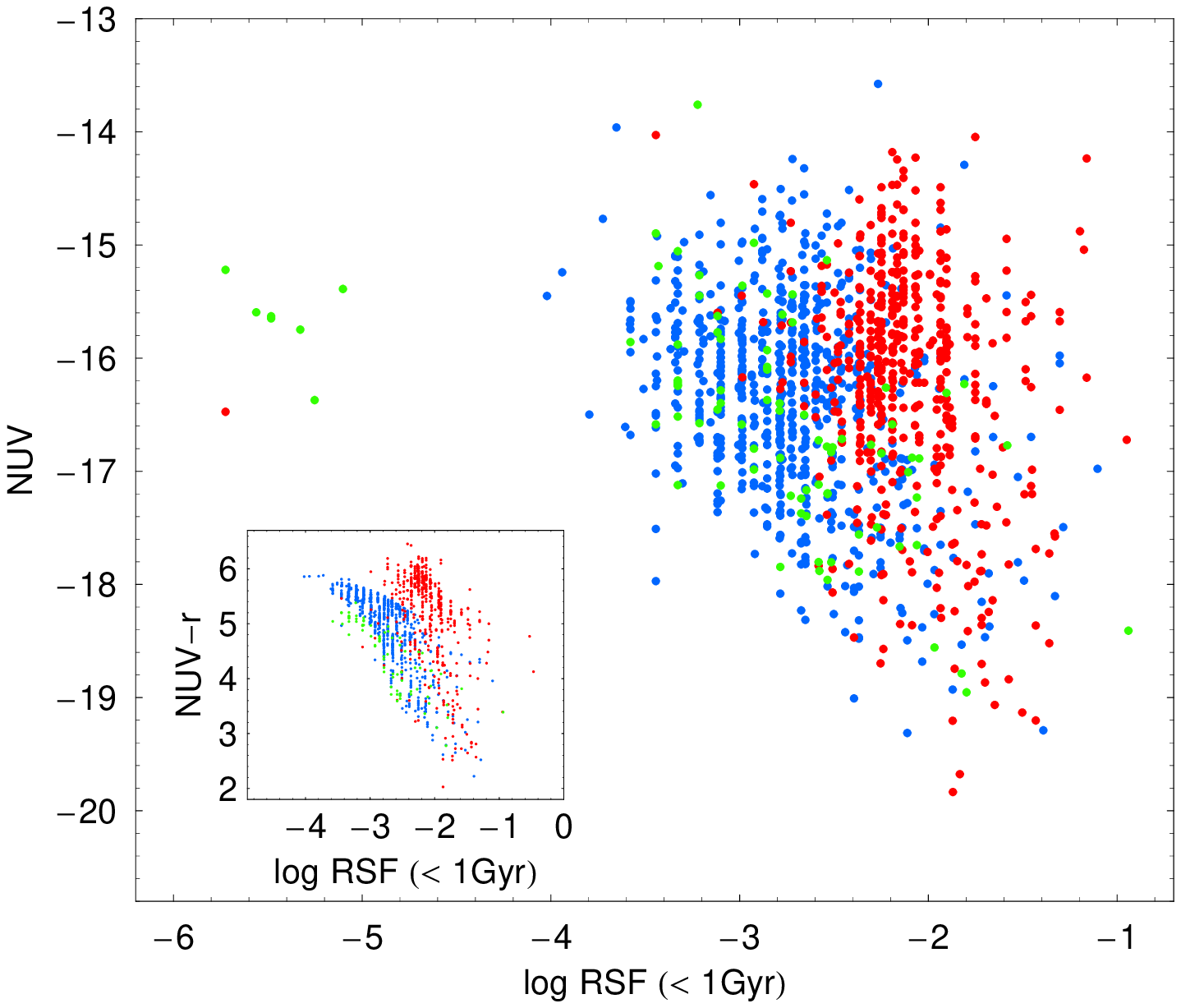, width=0.49\textwidth}\psfig{file=
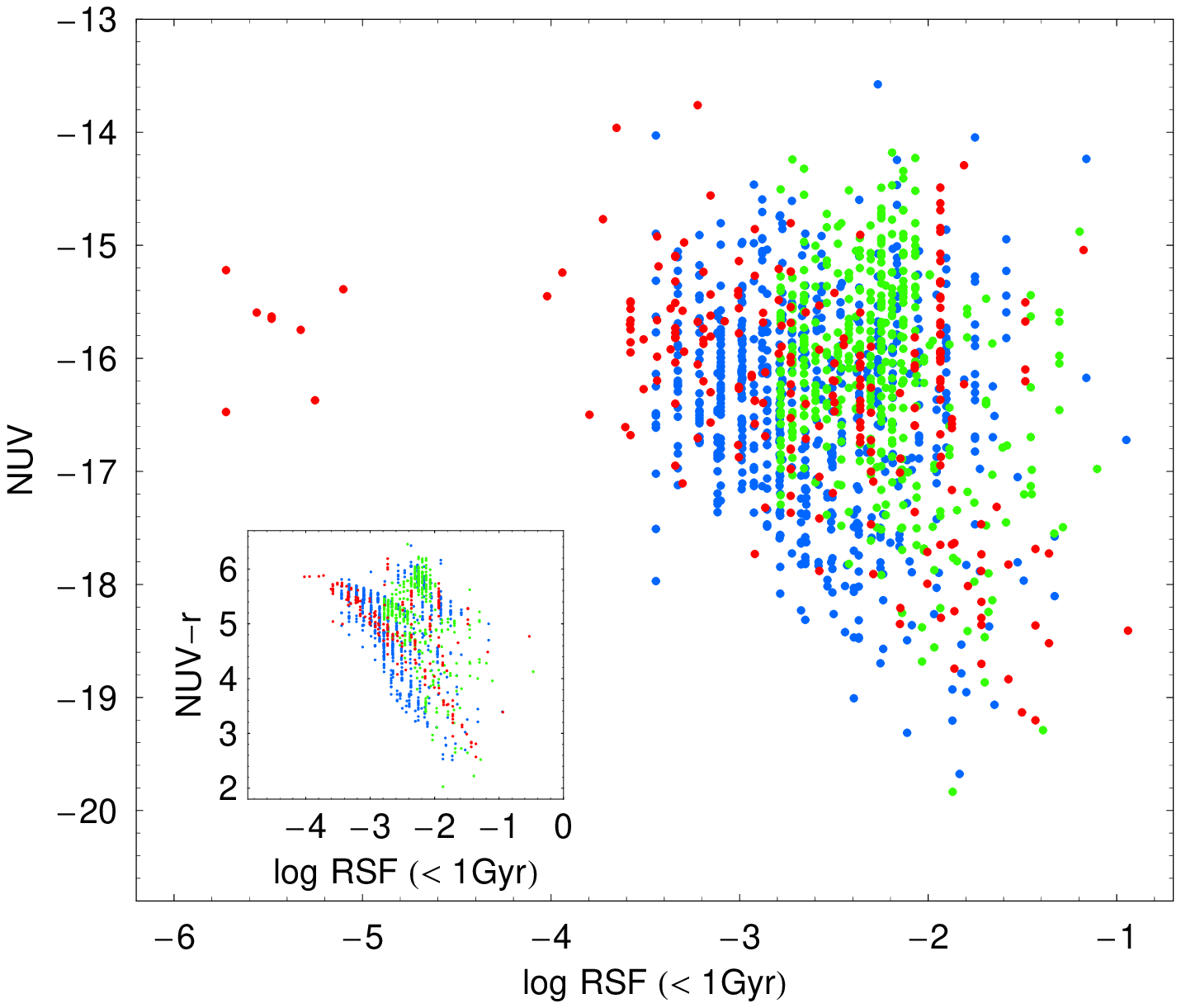, width=0.49\textwidth} \caption{NUV (and NUV-r)
vs RSF for a model leaving free all parameters. In the left panel
we colour for Z, with $Z=0.008$ (green), $Z=0.02$ (blue) and
$Z=0.05$ (red). In the right panel we colour for $\tau$ with $\tau
= 1 \, \rm Gyr$ (red), $\tau = 3 \, \rm Gyr$ (blue) and $\tau = 5
\, \rm Gyr$ (green).} \label{fig:RSF2}
\end{figure*}

Galaxies with a stronger ultraviolet flux are on average
characterized by  higher RSF, that translates into lower values of
$\tg - \tA$, while those with a lower NUV flux have SF that is
quenched early. This correlation is shown in Fig.
\ref{fig:tgtA_vs_NUVr_NUV}. In the inner panel, we show the trend
with colour NUV-r, implicit in Fig. \ref{fig:NUVr_vs_r_data}.

The results concerning the scale of SF are also interesting. When
considering the results obtained fitting our reference spectral
library, we find that only $\sim 15 \%$ of galaxies have $\tau =
1\, \rm Gyr$, while a more protracted SF is recovered for the
other galaxies in the sample, with $\sim 43 \%$ and $\sim 42 \%$
of these having $\tau = 3$ and $5\, \rm Gyr$. As a comparison, we
also fitted unperturbed SFs with $\tau \in (0.1,\,5) \, \rm Gyr$
and $Z \in (0.008,\,0.02,\,0.05)$ obtaining that $\sim 86 \%$ of
galaxies have $\tau \leq 1$, while less than $1\%$ have $\tau >
3\, \rm Gyr$ (see also \citealt{Tortora09}). Thus, the low SF
scales recovered when simple unperturbed exponential
SFs\footnote{Similar considerations can be made if we assume a
delayed SF.} are fitted to data can hide a galaxy population with
a more protracted background SF that is quenched in the late stage
of galaxy evolution for a feedback effect. These results are
consistent with the general scenario depicted for the color
evolution of E+A galaxies in \cite{Kaviraj07b}. They find that,
superimposed to an early burst of formation, a recent SF burst
(which typically takes place within $1 \, \rm Gyr$) over a
timescale ranging between 0.01 and 0.2 Gyrs is needed to match
galaxy colours. These galaxies are just migrating towards the red
sequence and show a SF quenching which is correlated with their
stellar mass and velocity dispersion and linked to different
sources of feedback (AGN or supernovae). However, although also
different models can have the same final result as our unperturbed
and truncated SFs for different class of galaxies, this work
reproduces typical timescales consistent with those discussed here
(see also Fig. \ref{fig:tscale}).

Fitting our reference library, we find that $\sim 35 \%$ of
galaxies have $\tg > t_{Univ}$, while this percentage decreases to
$\sim 15 \%$ for those galaxies with $\tg > t_{Univ} + 2$.
Excluding all  galaxies with $\tg
> t_{Univ}$ we find a median formation redshift of $z_{f} =
1.0^{+1.4}_{-0.4}$ (uncertainties are $25^{\rm th}$ and $75^{\rm
th}$ percentiles); note that the distribution has a strong tail
for high formation redshift. For \tA\ we find the best estimate
$z_{AGN} = 0.13 \pm 0.02$.

\subsection{Stellar mass fraction and AGN feedback}\label{sec:RSF}

Until now, we have discussed how the quenching event is linked to
colours and galaxy luminosities, but we have not expressly
discussed the amount of stellar mass that is produced. To
quantify this recent star formation, we define the RSF via the
amount of stars produced  (i.e. the produced stellar mass
fraction) in the last $1 \, \rm Gyr$ in the rest frame of the
galaxy (\citealt{Kaviraj07}, 2008). As discussed in the previous
subsections, this quantity is strictly dependent on the time of
the quenching event, and in particular on $\tg - \tA$. The latter
has been show to depend on colour NUV-r and the luminosity in the
NUV band, while it is less sensitive to the same quantities
obtained only using visible bands.

Galaxies with brighter ultraviolet fluxes and bluer NUV-r colours
have much more RSF with respect to fainter ones. This result is
not surprising, since an indication of it has been obtained by the
trends shown in Fig. \ref{fig:tgtA_vs_NUVr_NUV}. In Fig.
\ref{fig:RSF1} we show the RSF as a function of NUV and NUV-r. In
the left panel, we show the results using our reference library.
Obviously, only galaxies with $\tg - \tA$ less than $1 \, \rm Gyr$
survive in these plots and show the presence of RSF. In this
case, the trends are not so tight, since some galaxies depart
from them (red points). As a comparison, in the right panel we
show the results for a model with fixed values of $Z$ and $\tau$,
where the correlations are tighter. Coming back to the results we
obtained using our reference library, we show in Fig.
\ref{fig:RSF2} the same results already plotted in the left panel
of Fig. \ref{fig:RSF1}, but now coded for metallicity and $\tau$,
respectively in the left and right panels. These plots show that those
galaxies, which experience a significant RSF,
also have metallicities significant different from the average. In fact, galaxies with different metallicities
seem to stay on three different sequences, with a lot of
supersolar galaxies having red colours and a low NUV flux, but a
large SF. This trend is more evident if we look at the plot as a
function of colour. The trends with SF time-scale $\tau$ are less
clear, but some of those galaxies departing from the mean trend
have $\tau = 5 \, \rm Gyr$. Thus, some systematics might be
introduced into these results by the degeneracies between $Z$ and
$\tau$, and possibly with other parameters. However, our results
are robust, since qualitatively we recover some trends which do
not depend on the particular choice of spectral library.

For the model that leaves  all  of the parameters free, we find a
median RSF of $\sim 0.2^{+0.4}_{-0.2} \%$. In particular, galaxies
with the bluest colours (and with larger NUV fluxes) have RSF of
$\sim 1\%$, while the redder ones have RSF less than $0.1 \%$. On
the contrary, a lower median RSF of $\sim 0.1\%$ is observed for
the model with $\tau= 3 \, \rm Gyr$ and $Z= \Zsun$.

On average, our recovered RSF fractions are slightly lower than
those obtained in \cite{Kaviraj07}, but this is not surprising
since we follow a different approach. In detail, we adopt what we
can call a {\it blue-to-red} approach, since we have modelled
galaxies using an exponential SF. Thus, a galaxy is
initially blue,
and later it becomes red since SF is quenched. On the contrary,
using a {\it red-to-blue} approach, galaxies are modelled with a
single burst population, which predict systematically redder
colours than those obtained with a more complex SF. Thus, to match
the observations one needs a recent burst to make the colours
bluer. It is however worth to be noted that within scatter the two
analysis give fully compatible results.

Our results can also be connected with the downsizing scenario. In
fact, the fact that redder galaxies (i.e. with a lower NUV flux)
have less RSF with respect to the bluer ones can reproduce the
trends found in recent works (\citealt{Cowie96},
\citealt{Borch06}, \citealt{Bundy06}, \citealt{deLucia06},
\citealt{Trager00}, \citealt{Thomas05}, \citealt{Nelan05},
\citealt{Kaviraj08}, \citealt{Romeo+08}, \citealt{Tortora09}).
However, this connection has to be better analyzed and further
information could be recovered by extending the luminosity range.

\begin{figure*}
\psfig{file= 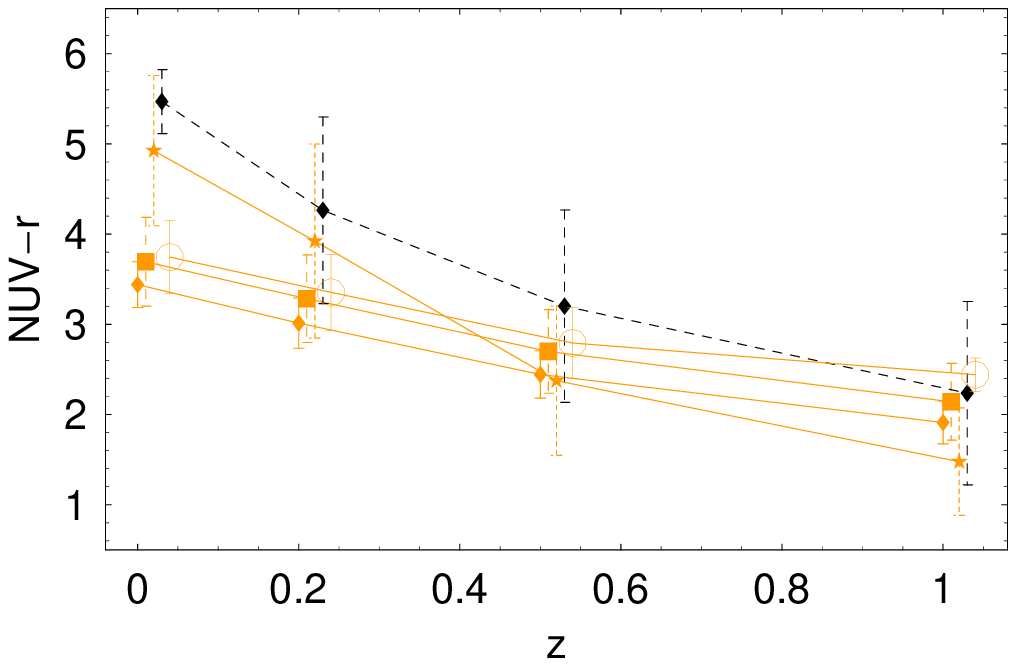, width=0.4\textwidth} \psfig{file=
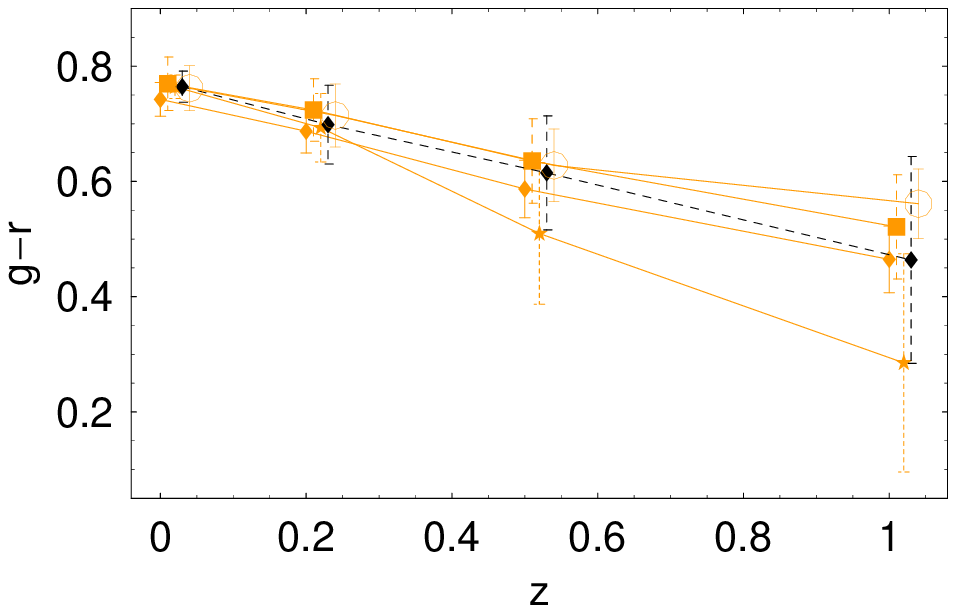,
width=0.4\textwidth}\\
\psfig{file= 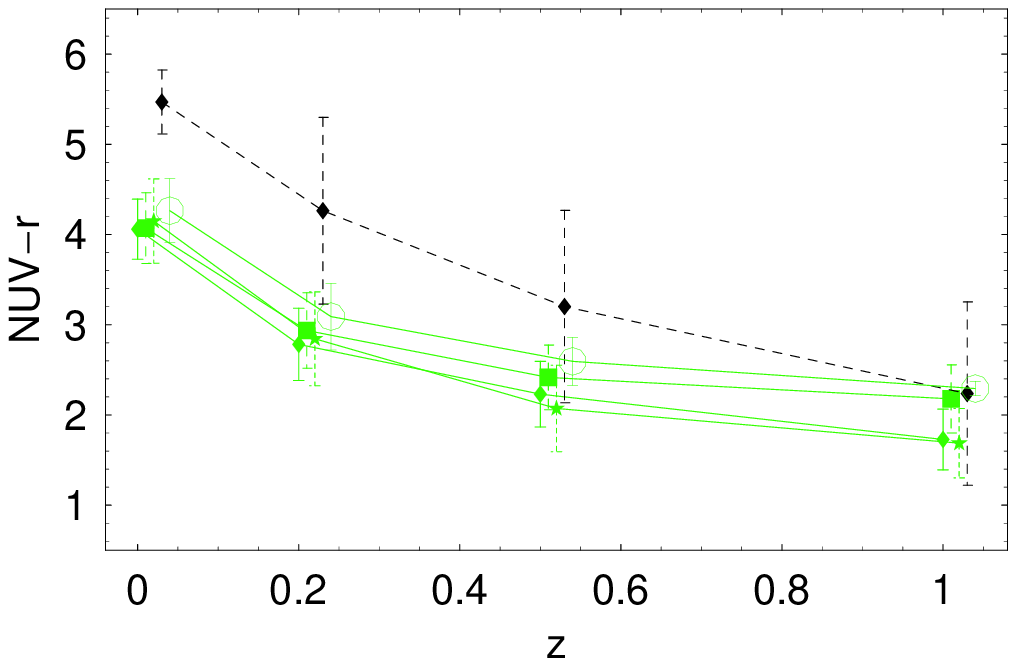, width=0.4\textwidth} \psfig{file=
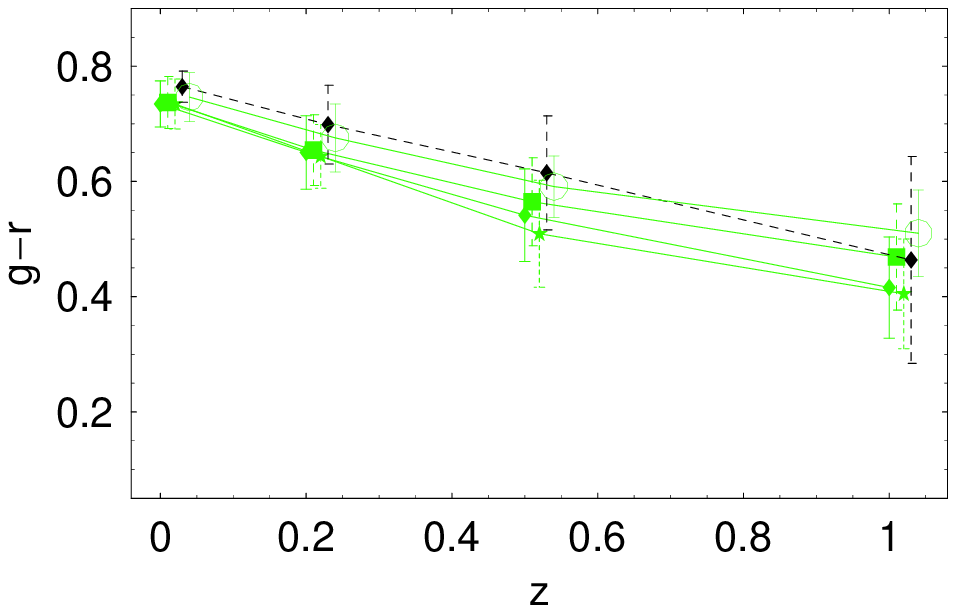,
width=0.4\textwidth}\\
\psfig{file= 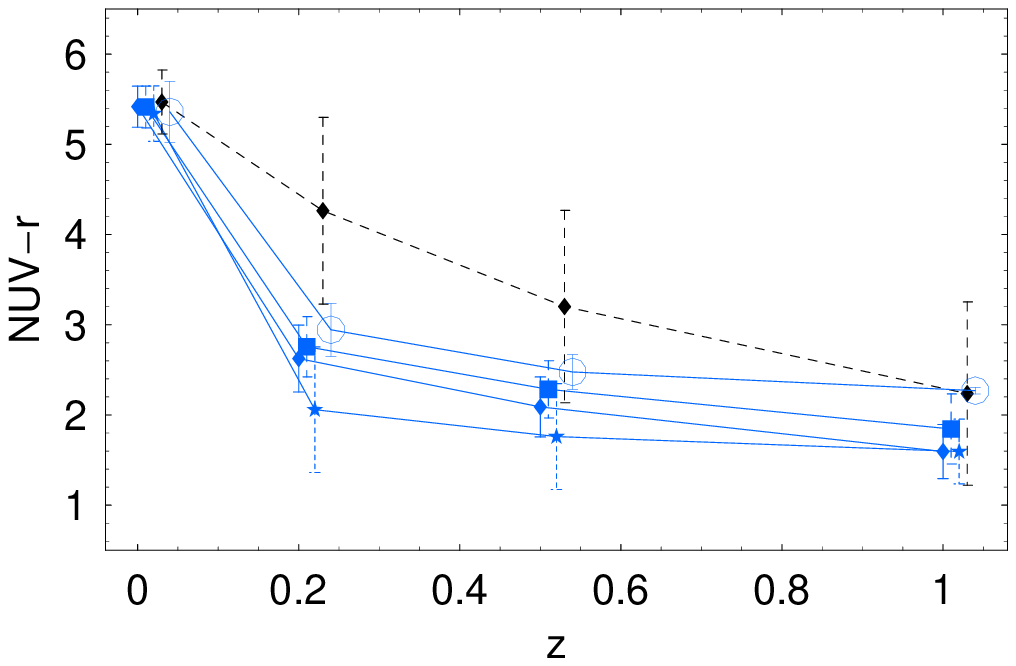, width=0.4\textwidth} \psfig{file=
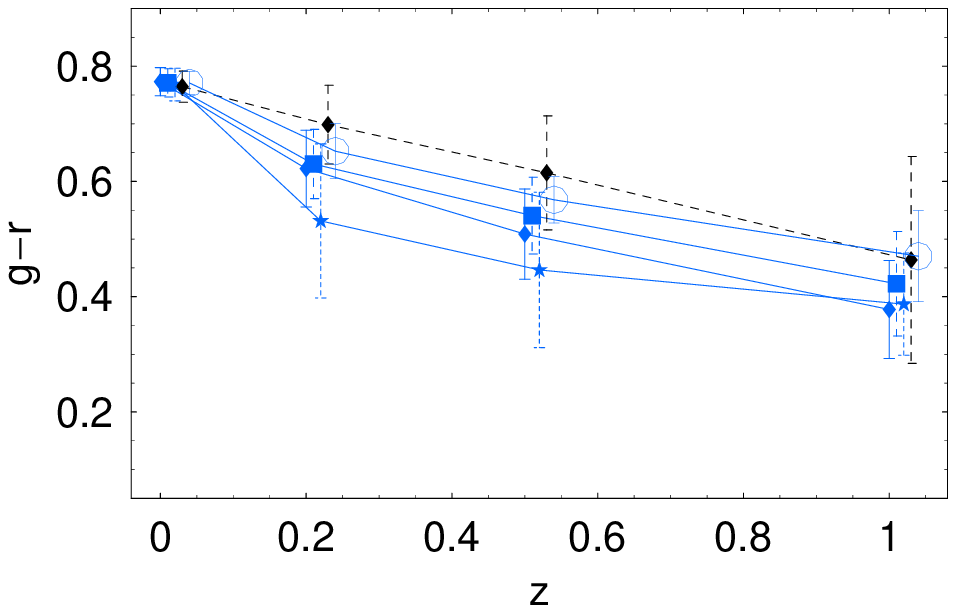,
width=0.4\textwidth}\\
\psfig{file= 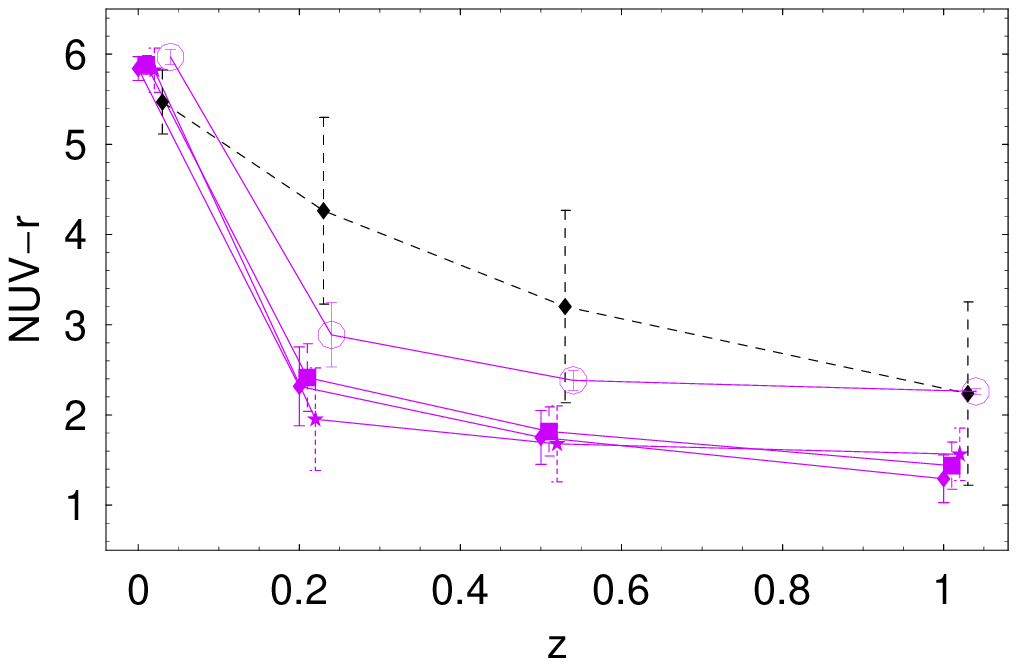, width=0.4\textwidth} \psfig{file=
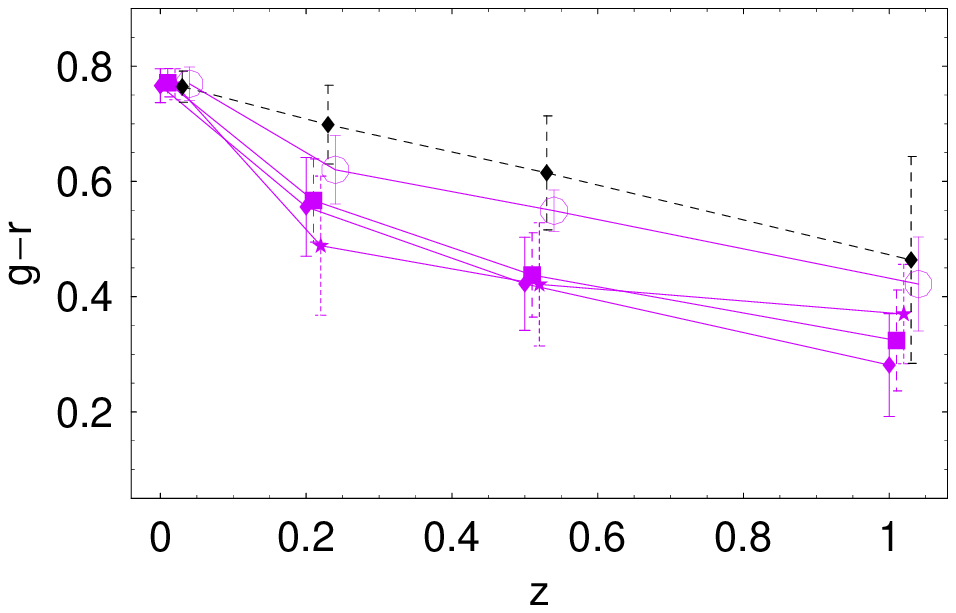,
width=0.4\textwidth}\\
\psfig{file= 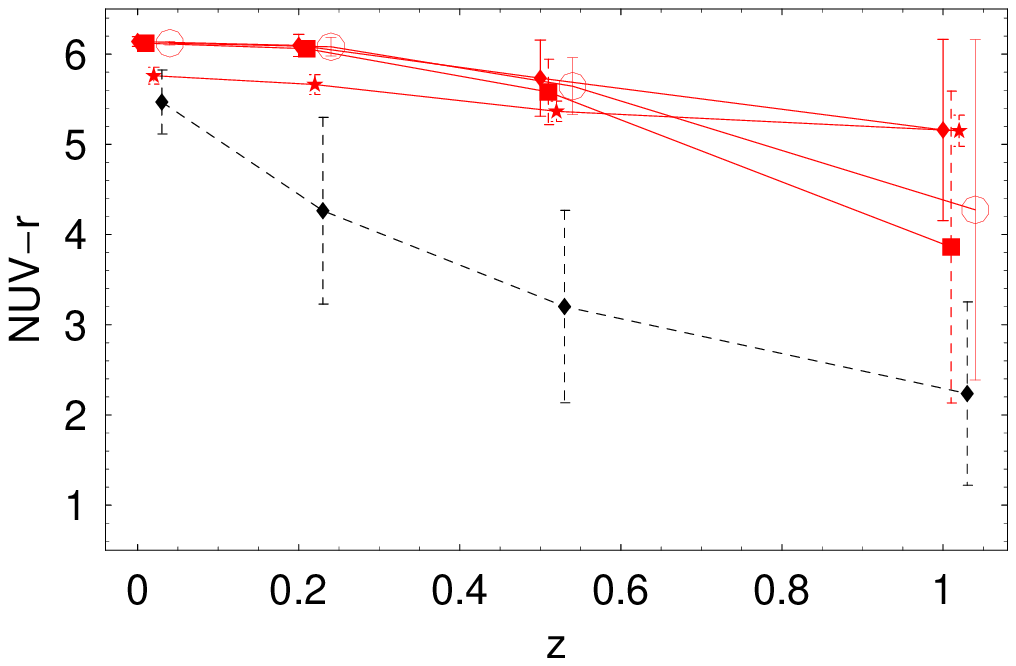, width=0.4\textwidth} \psfig{file=
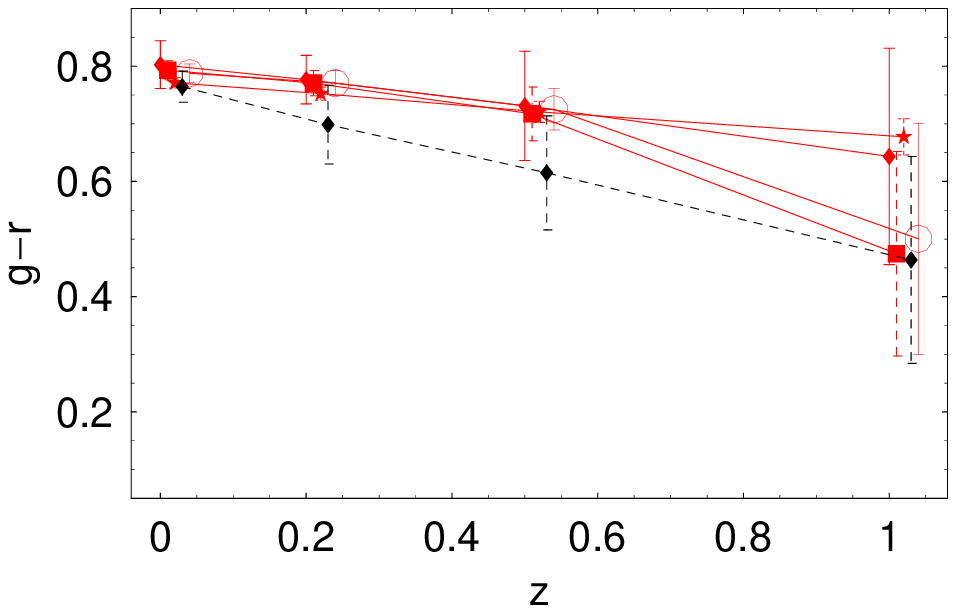,
width=0.4\textwidth}\\
\caption{Colour evolution predicted by fit. The different panels
refer to values of $\tg - \tA$ accordingly to the same colour
coding of Figs. \ref{fig:NUVr_vs_r_data},
\ref{fig:NUVr_vs_gr_data} and \ref{fig:RSF1}. We show the results
from our reference model leaving all parameters free to change
(star symbols and short-dashed error bars), $Z=\Zsun$ and $\tau =
3 \, \rm Gyr$ (diamond symbols and continuous error bars), two
combined models with a $Z=\Zsun$ burst superimposed to a truncated
exponential SF with $\tau =3 \, \rm Gyr$ with a free mass ratio
(box symbols and long-dashed error bars) and one with single burst
amounting to the $75\%$ of the total mass (circle symbols and thin
continuous error bars). The black symbols and bars are obtained
using an unperturbed SF with $\tau$ and $Z$ free to change
(averaged over all galaxies in the sample). At each redshift, the
different models are artificially shifted to make clearer the
appearance. {\it Left panels}. NUV-r vs z. {\it Right panels}. g-r
vs z. } \label{fig:Evolution}
\end{figure*}

\subsection{SED evolution}\label{sec:evolution}

In the previous subsection, we obtained results that indicate the
presence of a RSF that is connected with the quenching events.
Here, we quantify the evolutive path of galaxies in our
sample, discussing how colours and magnitudes (averaged over the
galaxy sample) evolve. A more detailed analysis of these results
is beyond the scope of this paper.

We verified that a large part of the galaxies in our sample
are not very sensitive to changes of the details of the spectral
library, since they predict a median $\tg - \tA$ that does not
change siginficatly (see Fig. \ref{fig:Distr_various_fits}). Thus, our
models are able to successfully describe the recent formation
history of galaxies, while the extrapolation to early phases of
galaxy evolution is influenced by the choice of unperturbed
synthetic spectra. However, this extrapolation can fail to
describe the earlier evolution, since galaxies which have experienced
various (minor or major) mergers, and AGNs can act to quench SF at
different epochs (\citealt{Khalatyan08}). Therefore, both SF,
magnitude and colour evolution can be very complex and is not
possible to efficiently probe them.

We map the evolution of rest-frame galaxy colours up to 7-8
billion of years ago, that we determine from our best fitted
models for 3 values of the look-back time $t_{lbt}$ corresponding
to redshift $z=0.2, \, 0.5, \, 1$. In Fig. \ref{fig:Evolution} we
show the median values and median deviations for $NUV-r$ and $g-r$
of galaxies in the sample. Actually, colour and luminosity
evolution is stronger for galaxies with a more recent event of SF
and quenching (i.e. lower $\tg\ - \tA$). These galaxies became red
only recently, while those with a quenching event that occurred
many Gyrs ago were already on the red sequence. While changes of
2-3 magnitudes in NUV-r are observed, g-r changes are $\lsim 1$
magnitude. In addition to our reference model, we show the
evolution of colours obtained using a library of truncated
SFs with $Z=\Zsun$ and $\tau = 3 \, \rm Gyr$ and two mixed models
consisting of a solar single burst and a truncated SF with
$Z=\Zsun$ and $\tau = 3 \, \rm Gyr$,  leaving free the proportion
of  the two populations, and fixing the single burst to 75\% of
the total mass. These two latter models leave the galaxy colours
for $\tg > \tA$ almost unchanged, while affecting the early
history of galaxies producing redder colours, this effect
depending on the ratio of the two populations. For comparison, we
show the median colours obtained fitting an unperturbed SF leaving
free to change $t_{gal}$, $Z$ and $\tau$.

These predicted tracks show an interesting result concerning the
strength of quenching, that is mainly evident in the NUV-r colour.
With the exception of galaxies with $\tg < \tA$ and those with an
early quenching event, the systems with an observed larger colour
difference (and thus on average large values of $\tg - \tA$) were
bluer and more star forming at $z \sim 0.2$. Thus, this result
seems to indicate that in galaxies with a large SF at $z \gsim \,
0.2$, the effect of AGN quenching can be stronger, to produce the
reddest galaxies we observe today. Nonetheless, recent X-ray and
optical selected analysis of high redshift AGNs
(\citealt{Martin07b}, \citealt{Nandra07}, \citealt{Silverman08a,
Silverman08b}) exhibit a large fraction of AGN in galaxies at
intermediate/blue colours, which could be driven by AGN feedback
toward red-sequence at $z \sim 0$.

The magnitudes NUV and r against redshift are plotted in Fig.
\ref{fig:Evolution2}. The estimated evolution  of the NUV
magnitude, being strongly sensitive to RSF,  can be different among
models, and is compared to those obtained using an unperturbed SF.
The NUV fluxes are systematically larger (corresponding to lower
$NUV-r$) than those estimated using unperturbed SFs. On the
contrary, the luminosity in the r band does not depend (or at
least depends only little) on a change in models and the
predictions agree quite well. This is not surprising, since redder
bands are less sensitive to recent or past SF.

\begin{figure*}
\psfig{file= 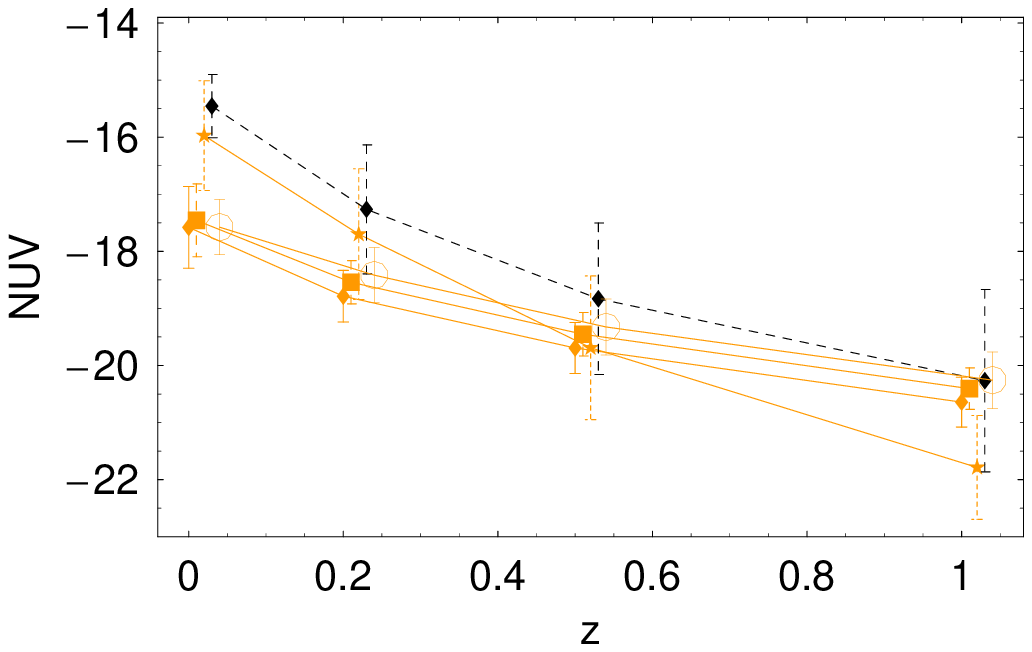, width=0.4\textwidth} \psfig{file=
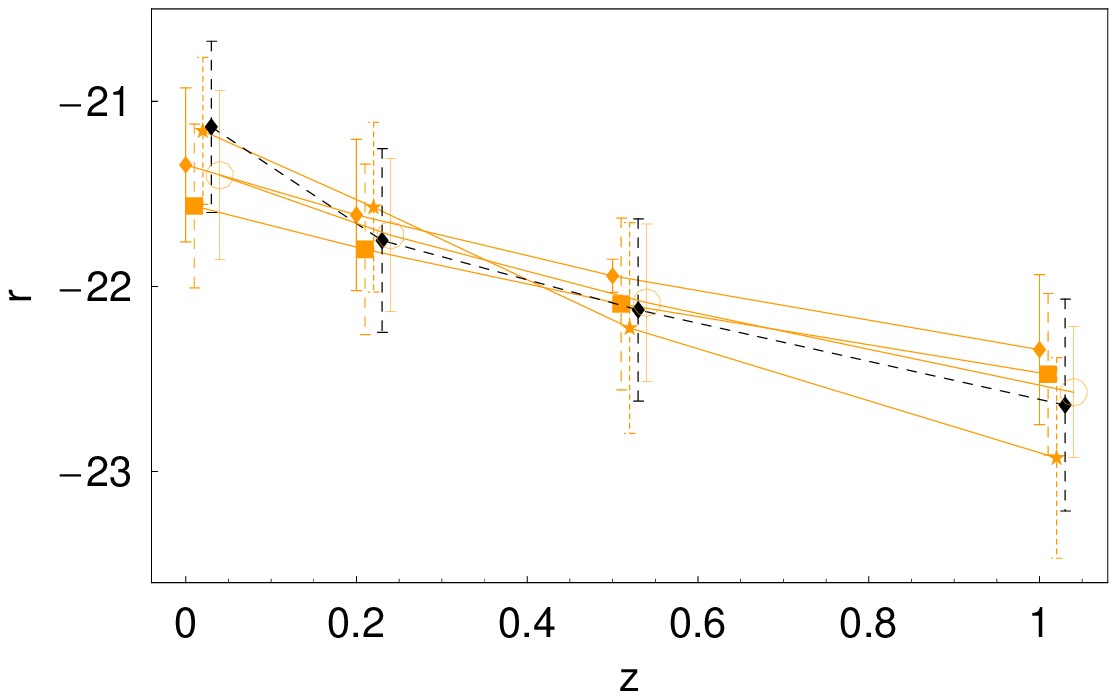, width=0.4\textwidth} \\
\psfig{file= 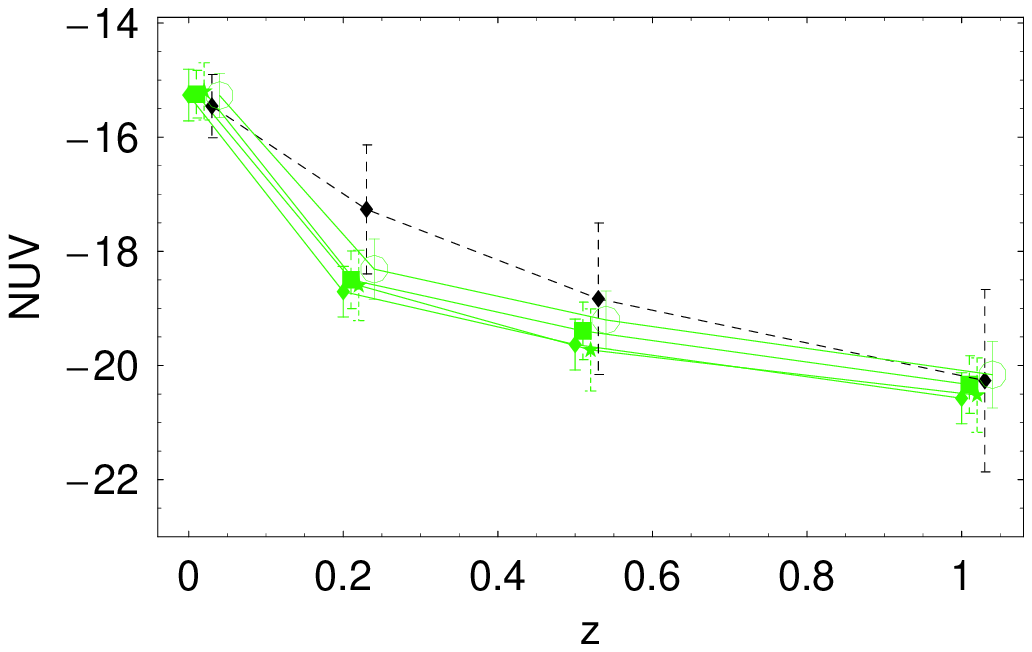, width=0.4\textwidth} \psfig{file=
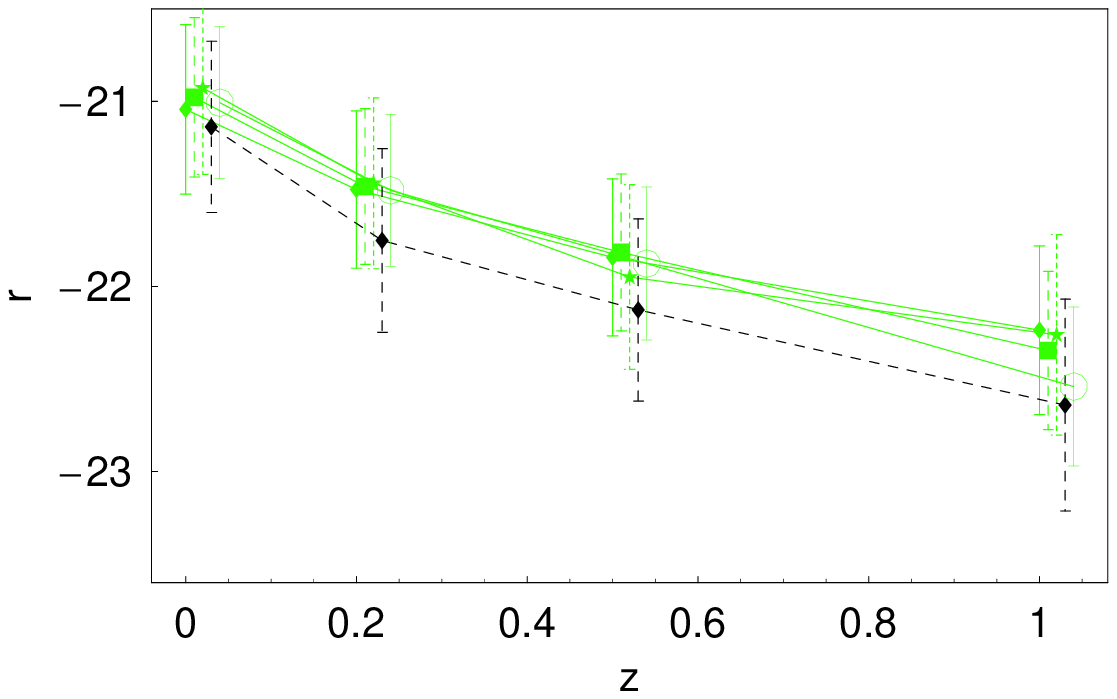, width=0.4\textwidth} \\
\psfig{file= 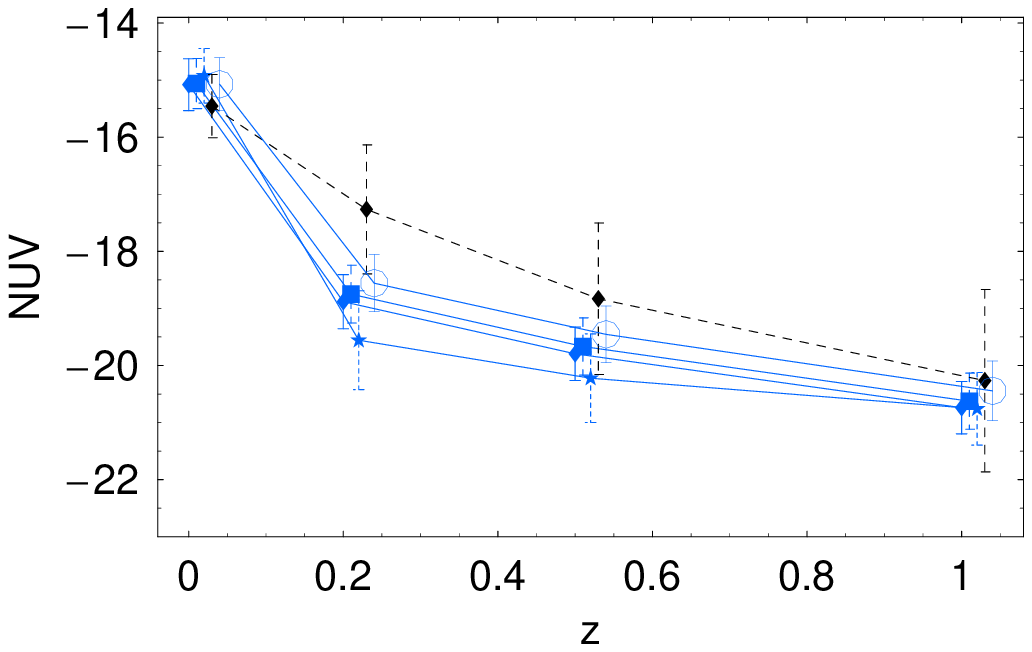, width=0.4\textwidth} \psfig{file=
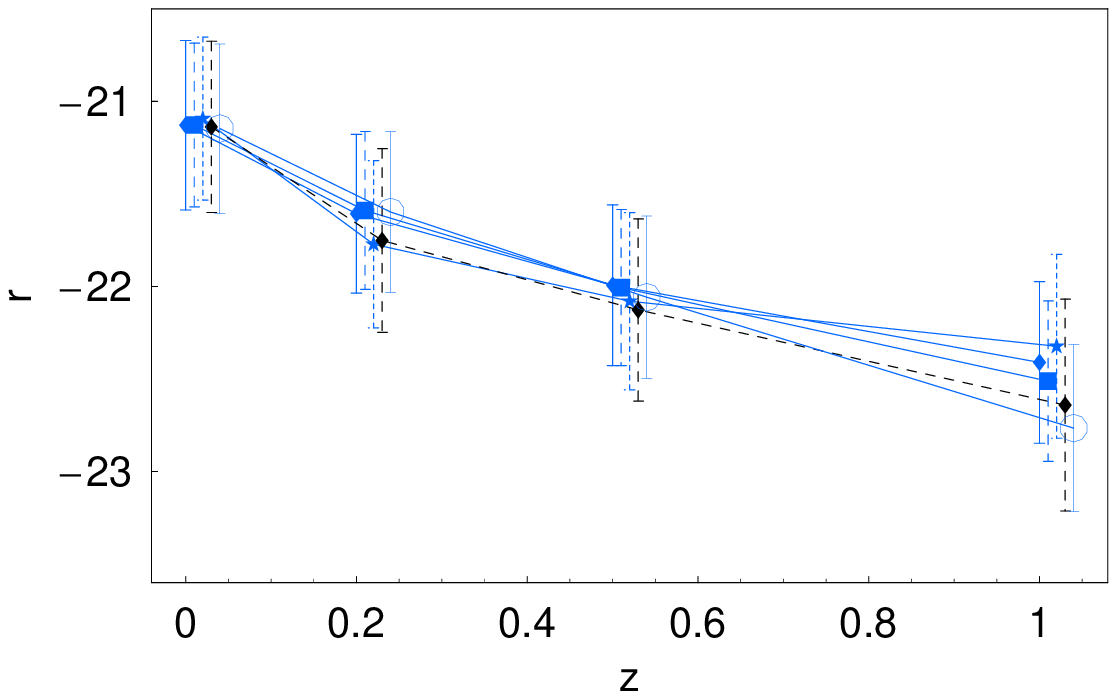, width=0.4\textwidth} \\
\psfig{file= 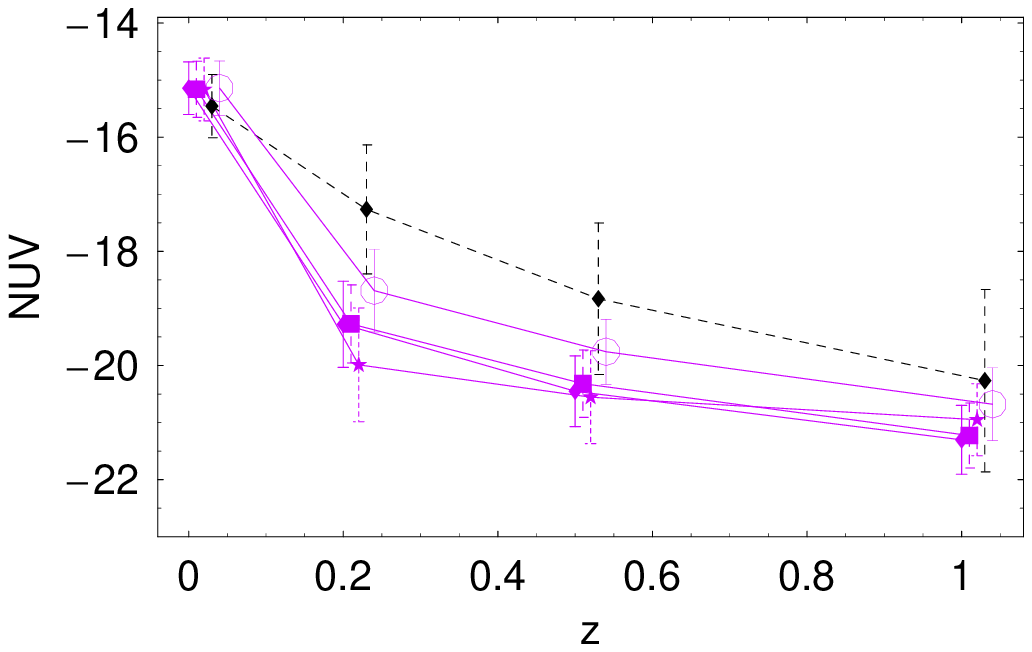, width=0.4\textwidth} \psfig{file=
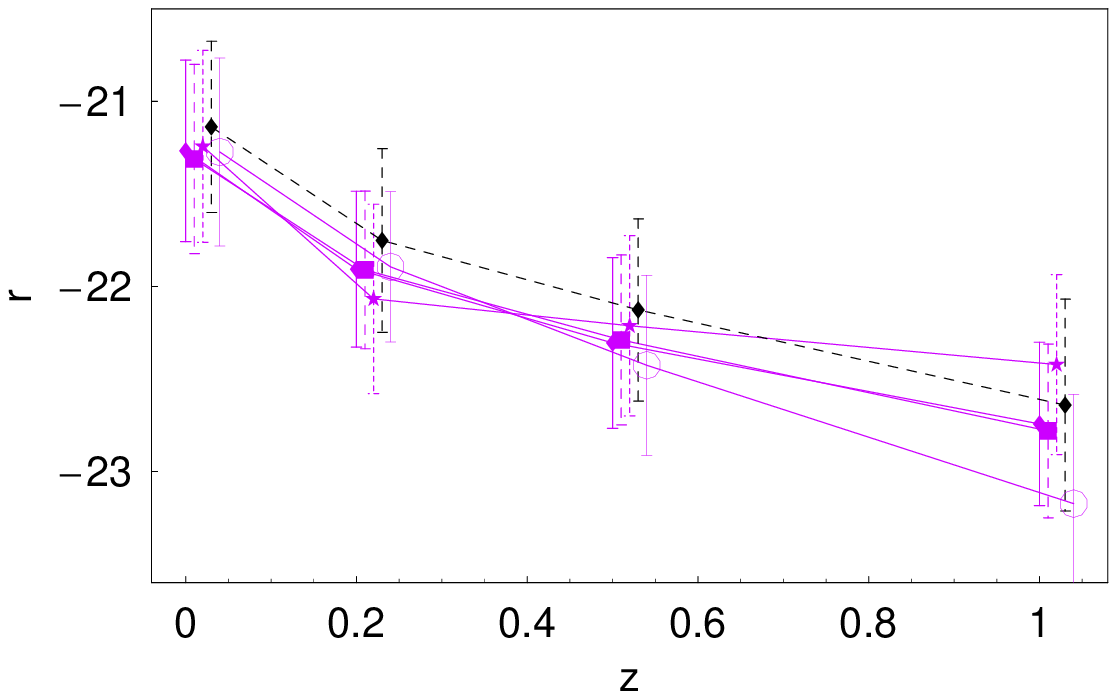, width=0.4\textwidth} \\
\psfig{file= 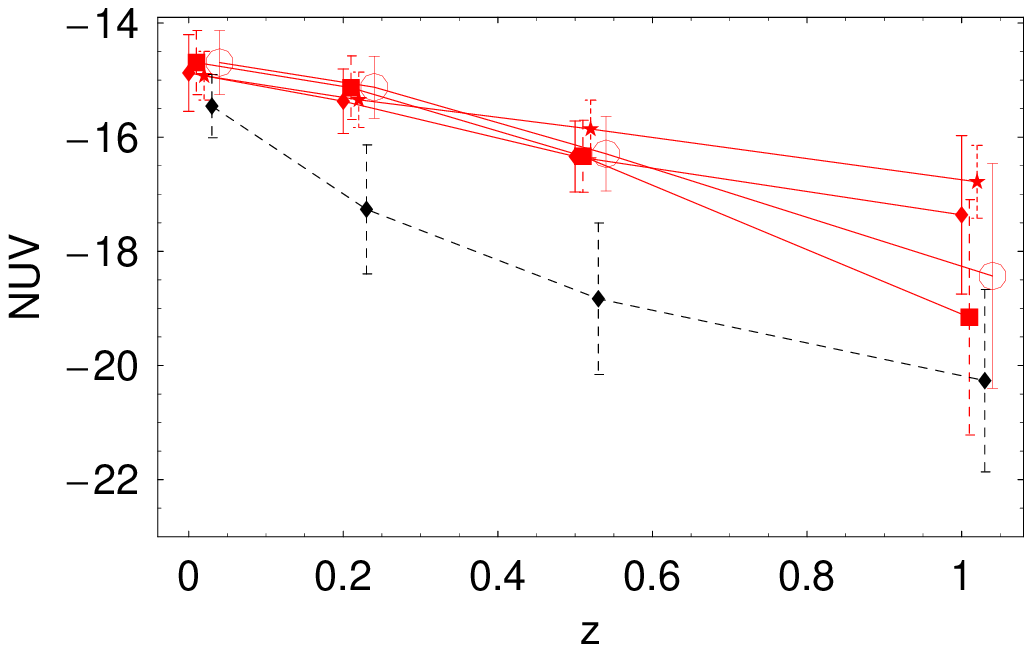, width=0.4\textwidth} \psfig{file=
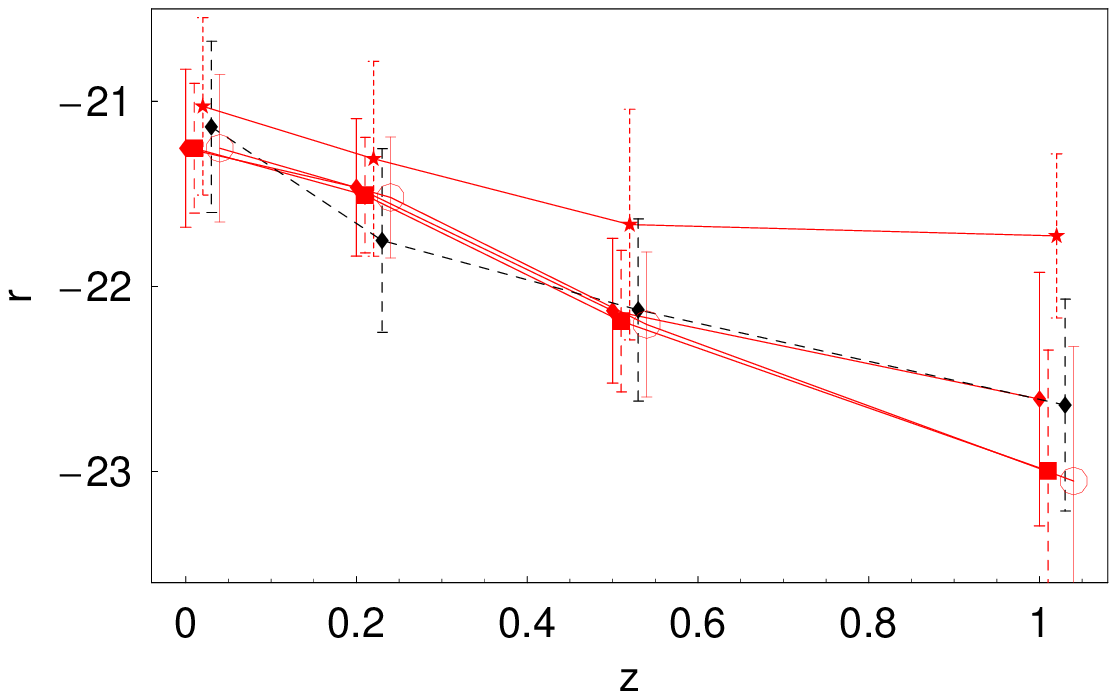, width=0.4\textwidth} \\
\caption{Magnitude evolution predicted by fit. See Fig.
\ref{fig:Evolution} for details about models and colour or symbol
code. {\it Left panels}. NUV vs z. {\it Right panels}. r vs z.}
\label{fig:Evolution2}
\end{figure*}

\section{Conclusions}\label{sec:conclusions}

In this work, we have attempted to build up a realistic model of AGN
feedback. We performed a hydrodynamical simulation to analyze the
impact of jets produced by AGNs on a inhomogeneous medium,
properly representing cold gas clouds that form stars in galaxies.
In our simulation, which extends and generalizes a set that we
previously performed (\citealt{AS08}), a powerful jet $P_{j} =
10^{46} \, \rm erg \, s^{-1}$ propagates within an inhomogeneous,
two-phase ISM, containing a realistic distribution of star-forming
clouds. SFRs in clouds are described using the empirical SK
prescription, where the SFR depends on the mass density of the
star forming regions of the clouds. In an early phase, the shocks
advancing before the expanding cocoon tend to compress the cold
clouds, without significantly changing the temperature of the
medium,  thus increasing the SF. Later, when the cocoon has
propagated within the medium, the temperature of both the medium
and of the clouds increases significantly and the mass of clouds
is reduced, also due to KH instabilities. Thus, at the beginning,
a positive feedback increases SF, but the dominant effect is the
negative feedback that quenches SF in a time of $\sim 2-3 \times
10^{6} \, \rm yr$. One interesting result of this paper is that for
the first time a hydrodynamical simulation allows the determination of the
effect of jets emitted by AGN on SF in galaxies.
Previous work has insted relied relying on empirical prescriptions to take into
account AGN feedback (\citealt{Granato01, Granato04},
\citealt{Cattaneo06}, \citealt{Martin07}), and are thus more robustly
supported by our results.

The high jet power in this simulation is probably the main reason
of the fact that, at the end of the simulation, all the gas
within the computational volume is in a physical state for which
all SF is quenched. For lower injection powers, the cocoon
expansion will be halted before, and the region where SF is
quenched will consequently be smaller.

Based on the results of this simulation, we develop a more general
model which assumes a SFR for ETGs composed of a background SF
with a more or less extended duration, that is quenched by a
feedback effect such as that analyzed in the first part of this
paper. However, we stress here that this general prescription
can also describe the effect of other quenching mechanisms, like
merging, harassment, etc. Restricting ourselves to the case of a
high power jet, as in the present paper, the typical timescales of
quenching are very small, and no other sources of feedback can stop
SF within such a short time. Note also that galaxy merging seems
to be one of the mechanisms which activate AGNs, which after the
burst induced by the merger is able to quench the residual SF in
the remnant (\citealt{DiMatteo05}, \citealt{Khalatyan08}).

Our paradigm is that luminous ETGs are an exception within the zoo
of galaxies in the universe. We suppose that the normality in
galaxy samples is represented by fainter ETGs, that have a
protracted SF, but by the effect of feedback they are quenched and
colours redden. This could be one of the missing physical
ingredients which could explain the short duration of SF, when
exponential SF are fitted to data (e.g., \citealt{Tortora09}). On
the contrary, semi-analytical galaxy simulations predict more
protracted SFR (\citealt{deLucia06}): supernovae feedback was
taken into account in these simulations, but the disagreement with
other observations can be due to an improperly accounted for
contribution from AGNs. \cite{Springel05} simulate the effect of
AGNs in the merging of two massive and gas-rich spiral galaxies,
obtaining similar results. They found that SF is inhibited with
respect to the case without black holes; interestingly, without
AGNs this kind of merger is not able to produce red ETGs, that
remain blue (with residual SF) even after several Gyrs. On the
contrary, the presence of black holes redden galaxy colours much
faster, giving  $u-r \sim 2.3$ in less than 1 Gyr after the
beginning of the merging process.

We assume an unperturbed SF law with a scale $\tau = 1, \,3, \, 5
\, \rm Gyr$, that is quenched at \tA\, leaving free to change
these parameters as well as  galaxy ages and metallicities. We fit
these models to observations from a cross-matched SDSS+GALEX
catalog. Confirming the results in \cite{Kaviraj07},  the UV has
been shown to be a strong indicator of SF, in particular, the
UV/optical colour $NUV-r$ allows us  to select galaxies with
different levels of SF, stopped in the last Gyrs or still in
action. We show that the quantity $\tg - \tA$ is able to describe
the physical state of galaxies, indicating how much time ago SF is
stopped. The largest number of galaxies have $\tg - \tA \sim 0.5 -
1 \, \rm Gyr$, indicating the necessity of a RSF phenomenon (until
$1-2 \, \rm Gyr$ ago), that is quenched by AGNs. Galaxies with
larger $NUV-r$ have higher values of $\tg - \tA$, i.e. SF is
quenched early in their history, while lower values of $NUV-r$
correspond to galaxies with a more recently quenched SF. Finally,
galaxies with a SF not affected by AGN have the flattest  $NUV-r$
colours. This distinction is not so clear (or absent) for visible
colours. These results are shown in Figs. \ref{fig:NUVr_vs_r_data}
and \ref{fig:NUVr_vs_gr_data}. The epoch of the quenching event
appears to be correlated with ultraviolet flux: galaxies bluer and brighter
in NUV band are also those with a more recent (or
absent) quenching event. As shown in Fig.
\ref{fig:Distr_various_fits}, $\tg - \tA$ is found to negligibly
depend on the details of background spectral library (e.g., $\tau$
and $Z$), at least for the bulk of galaxies: thus we obtain a
robust estimate of the quenching time and RSF. Our analysis is
less sensitive to the earliest phases of galaxy history,
notwithstanding the mean evolution of galaxy population here
analyzed, as shown in Figs. \ref{fig:Evolution} and
\ref{fig:Evolution2}. Finally, a shortcoming of this analysis is
related to the timescale of quenching event, since softer power
jets would inhibit SF within a larger timescale. Such a slower
quenching model would be fitted to the colours by a larger $\tg -
\tA$, thus our results for this parameter have to be interpreted
as lower limits.

One of the most significant results we have found is that the
NUV-g colour index is very sensitive to the presence of a very
young stellar population. Typical enhancement of 2-3 magnitudes of
NUV-g are observed, with regard to the no-feedback case. Our
findings agree with some current simulations that invoke two
different modes of AGN feedback. A so called `quasar mode' assumes
that during a major merger event at high redshift, a fraction of
the gas accreted by a central black hole is injected into the gas
of the host galaxy, quenching SF (\citealt{Springel05,
Springel05b}, \citealt{DiMatteo05}). At later times, another
effect is important, i.e. the `radio mode', that is responsible
for making galaxies quiescent,  and this effect is driven by
low-level AGNs (\citealt{Croton06}, \citealt{Bower06}). More
recently, \cite{Schawinski08} speak of  a `truncation mode' to
indicate AGN feedback at high redshift. At recent epochs, no such
strong activities or powerful radio jets have been observed, and
so this kind of AGN feedback is often referred to as the
`suppression mode'. These authors suggest that this process could
leave a residual SF (e.g. \citealt{Schawinski06}), but is able to
move galaxies along the red sequence.

Further refinements of our analysis are needed. Firstly, it is
important to enlarge the sample of galaxies and analyze a more
extended range of magnitudes. The galaxies under analysis, due to
constraints on magnitude, are relatively bright with $M_{B} \lsim
-19$. Different kinds of analysis seem to indicate that brighter
and more massive ETGs ($M_{B} \lsim -20.5$ and $M_{\star} \gsim \,
10^{11} \, \rm \Msun$) are fundamentally different from fainter
and less massive ones ($M_{B} \gsim -20.5$ and $M_{\star} \lsim \,
10^{11} \, \rm \Msun$). In these two different luminosity and mass
regimes, the size-luminosity or size-mass (\citealt{Shen2003}) and
Faber-Jackson relations (\citealt{MG05}) have different slopes.
Also, the  Sersic index is changing with luminosity
(\citealt{PS97}), and dark matter is shown to have a bivariate
behaviour in the two ranges (\citealt{Tortora09}), suggesting that
physical phenomena allowing the formation of galaxies of different
mass and luminosity can be variegate. Thus, a future analysis
should also be directed at studying a wider range of luminosities
and masses to map the transition from the blue cloud to the
red sequence. Also to connect our results with the downsizing
scenario, we need to enlarge the sample to fainter magnitudes,
since strong changes in observable quantities such as galaxy age,
\tA\, $\tau$ and $Z$ are probably visible in these luminosity
regimes. We are planning to do other simulations changing the main
input parameters, in order to have a more reasonable model of AGN
feedback which depends, e.g., on the jet power, linking it to main
galaxy observables (\citealt{2006ApJ...637..669L}). Comparison
with simulations will allow to quantify how much AGN feedback has
to be implemented, to make both semi-analytical and hydrodynamical
simulations able to correctly predict the main properties of
galaxies we observe. Such a more complex model of quenching
would be parameterized as a function of the main parameters of the
system and compared to other AGN feedback prescriptions
(\citealt{Granato01, Granato04}, \citealt{Cattaneo06},
\citealt{Martin07}).

Linking observations at low redshift, such as  those analyzed
here, with high redshift data (\citealt{Martin07b},
\citealt{Nandra07}, \citealt{Silverman08a, Silverman08b}) could
certainly be a powerful way to shed light on the galaxy evolution
scenario, and in particular the role that AGNs have in the early
and later phases of the SF history of galaxies.

\section*{Acknowledgments}

We thank the anonymous referee for his report that have helped us
to improve the paper. The work of V.A.-D. has been supported by
the European Commission, under the VI Framework Program for
Research \& Development, Action ``{\em Transfer of Knowledge}''
contract MTKD-CT-002995 (``{\em Cosmology and Computational
Astrophysics at Catania Astrophysical Observatory}''), and by the
EC-funded project HPC-Europa++, grant HPC-Europa no. 1127. V.A.-D.
would also express his gratitude to the staff of the Institute of
Theoretical Astrophysics, University of Heidelberg Germany, of the
subdepartment of Astrophysics, Department of Physics, University
of Oxford, and of the Beecroft Institute for Particle
Astrophysics, for the kind hospitality during
the completion of this work.\\
The software used in this
work was partly developed by the
DOE-supported ASC/Alliance Center for Astrophysical Thermonuclear
Flashes at the University of Chicago. Finally,
this work makes use of results produced by the PI2S2 Project managed
by the Consorzio COMETA, a project co-funded by the Italian Ministry
of University and Research (MIUR) within the {\em Piano Operativo Nazionale
"Ricerca Scientifica, Sviluppo Tecnologico, Alta Formazione" (PON
2000-2006)}. More information is available at http://www.pi2s2.it (in
italian) and http://www.trigrid.it/pbeng/engindex.php\\
Funding for the SDSS and SDSS-II has been provided by the Alfred
P. Sloan Foundation, the Participating Institutions, the National
Science Foundation, the U.S. Department of Energy, the National
Aeronautics and Space Administration, the Japanese Monbukagakusho,
the Max Planck Society, and the Higher Education Funding Council
for England. The SDSS Web Site is http://www.sdss.org/.

The SDSS is managed by the Astrophysical Research Consortium for the Participating Institutions.
The Participating Institutions are the American Museum of Natural History, Astrophysical Institute Potsdam,
University of Basel, University of Cambridge, Case Western Reserve University, University of Chicago, Drexel
University, Fermilab, the Institute for Advanced Study, the Japan Participation Group, Johns Hopkins University,
the Joint Institute for Nuclear Astrophysics, the Kavli Institute for Particle Astrophysics and Cosmology,
the Korean Scientist Group, the Chinese Academy of Sciences (LAMOST), Los Alamos National Laboratory,
the Max-Planck-Institute for Astronomy (MPIA), the Max-Planck-Institute for Astrophysics (MPA),
New Mexico State University, Ohio State University, University of Pittsburgh, University of Portsmouth,
Princeton University, the United States Naval Observatory, and the University of Washington.\\

\appendix

\end{document}